\documentclass[12pt,preprint]{aastex}
\usepackage{natbib}
\RequirePackage{lineno}
\usepackage{multirow}
\usepackage{epsfig}
\usepackage{longtable}
\usepackage{amsmath,amssymb}

\shorttitle{IceCube 40-string Neutrino Search}
\shortauthors{R.~Abbasi et al.}

\begin{document}


\title{Time-Integrated Searches for Point-like Sources of Neutrinos with the 40-String IceCube Detector}


\author{
IceCube Collaboration:
R.~Abbasi\altaffilmark{1},
Y.~Abdou\altaffilmark{2},
T.~Abu-Zayyad\altaffilmark{3},
J.~Adams\altaffilmark{4},
J.~A.~Aguilar\altaffilmark{1},
M.~Ahlers\altaffilmark{5},
K.~Andeen\altaffilmark{1},
J.~Auffenberg\altaffilmark{6},
X.~Bai\altaffilmark{7},
M.~Baker\altaffilmark{1}
% ...
}
\altaffiltext{1}{Dept.~of Physics, University of Wisconsin, Madison, WI 53706, USA}
\altaffiltext{2}{Dept.~of Subatomic and Radiation Physics, University of Gent, B-9000 Gent, Belgium}
\altaffiltext{3}{Dept.~of Physics, University of Wisconsin, River Falls, WI 54022, USA}
\altaffiltext{4}{Dept.~of Physics and Astronomy, University of Canterbury, Private Bag 4800, Christchurch, New Zealand}
\altaffiltext{5}{Dept.~of Physics, University of Oxford, 1 Keble Road, Oxford OX1 3NP, UK}
\altaffiltext{6}{Dept.~of Physics, University of Wuppertal, D-42119 Wuppertal, Germany}
\altaffiltext{7}{Bartol Research Institute and Department of Physics and Astronomy, University of Delaware, Newark, DE 19716, USA}
% ...



\begin{abstract}

We present the results of time-integrated searches for astrophysical neutrino sources in both the northern and southern skies. 
Data were collected using the partially-completed IceCube detector in the 40-string configuration recorded between 2008 April 5 and 2009 May 20, totaling 375.5 days livetime.
An unbinned maximum likelihood ratio method is used to search for astrophysical signals.  The data sample contains 36,900 events: 14,121 from the northern sky, mostly muons induced by atmospheric neutrinos and 22,779 from the southern sky, mostly high energy atmospheric muons.  The analysis includes searches for individual point sources and  targeted searches for specific stacked source classes and spatially extended sources.
While this analysis is sensitive to TeV--PeV energy neutrinos in the northern sky, it is primarily sensitive to neutrinos with energy greater than about 1~PeV in the southern sky.  
No evidence for a signal is found in any of the searches.
Limits are set for neutrino fluxes from astrophysical sources over the entire sky and compared to predictions.  
The sensitivity is at least a factor of two better than previous searches (depending on declination), with 90\% confidence level muon neutrino flux upper limits being between $E^{2} dN/dE \sim 2-200 \times 10^{-12}~\mathrm{TeV \, cm^{-2} \, s^{-1}}$ in the northern sky and between $3-700 \times 10^{-12}~\mathrm{TeV \, cm^{-2}\, s^{-1}}$ in the southern sky.  
The stacked source searches provide the best limits to specific source classes.  
The full IceCube detector is expected to improve the sensitivity to $E^{-2}$ sources by another factor of two in the first year of operation.

\end{abstract}




\section{Introduction}
\label{sec1} 


Neutrino astronomy is tightly connected to cosmic ray (CR) and gamma ray astronomy, since neutrinos likely share their origins with these other messengers.
With a possible exception at the highest observed energies, CRs propagate diffusively losing directional information due to magnetic fields, and both CRs and gamma rays at high energies are absorbed due to interactions on photon backgrounds.
Neutrinos, on the other hand, are practically unabsorbed {\it en route} and travel directly from cosmological sources to the Earth.  
Neutrinos are therefore fundamental to understanding CR acceleration processes up to the highest energies, and the detection of astrophysical neutrino sources could unveil the origins of hadronic CR acceleration.  
Whether or not gamma ray energy spectra above about 10~TeV can be accounted for by only Inverse Compton processes is still an open question.  
Some observations suggest contributions from hadronic acceleration processes \citep{Morlino:2009ci,boettcher2009}.
Acceleration of CRs is thought to take place in shocks in supernova remnants (SNRs) or in jets produced in the vicinity of accretion disks by processes which are not fully understood. 
Black holes in active galactic nuclei (AGN), galactic micro-quasars and magnetars, or disruptive phenomena such as collapsing stars or binary mergers leading to gamma ray bursts (GRBs), all characterized by relativistic outflows, could also be powerful accelerators.  
The canonical model for acceleration of CRs is the Fermi model \citep{PhysRev.75.1169}, called first-order Fermi acceleration when applied to non-relativistic shock fronts.  
This model naturally gives a CR energy spectrum with spectral index similar to $E^{-2}$ at the source.  
The neutrinos, originating in CR interactions near the source, are expected to follow a similar energy spectrum.  
More recently, models such as those in \citet{Caprioli:2009fv} can yield significantly harder source spectra.
In the framework of these models it is possible to account for galactic CR acceleration to energies up to the knee, at about $Z \times 4\times10^{15}$~eV, where $Z$ is the atomic number of the CR.  
Extragalactic sources, on the other hand, are believed to be responsible for ultra-high energy CRs observed up to about $10^{20}$~eV.

The concept of a neutrino telescope as a 3-dimensional matrix of photomultiplier tubes (PMTs) was originally proposed by \citet{Markov1961385}.
These sensors detect the Cherenkov light induced by relativistic charged particles passing through a transparent and dark medium such as deep water or the Antarctic ice sheet.  
The depth of these detectors helps to filter out the large number of atmospheric muons, making it possible to detect the rarer neutrino events.  
The direction and energy of particles are reconstructed using the arrival time and number of the Cherenkov photons.  
High energy muon neutrino interactions produce muons that can travel many kilometers and are almost collinear to the neutrinos above a few TeV.  
The first cubic-kilometer neutrino telescope, IceCube, is being completed at the South Pole.  
IceCube has a large target mass.  
This gives it excellent sensitivity to astrophysical neutrinos, enabling it to test many theoretical predictions.

Reviews on neutrino sources and telescopes can be found in \citet{Anchordoqui:2009nf, Chiarusi:2009ng, Becker:2007sv, Lipari:2006uw, Bednarek:2004ky, Halzen:2002pg, Learned:2000sw, Gaisser:1994yf}.  Recent results on searches for neutrino sources have been published by IceCube in the 22-string configuration \citep{Abbasi:2009cv,Abbasi:2009iv}, AMANDA-II \citep{AMANDA7YR_Collaboration:2008ih}, Super-Kamiokande \citep{Thrane:2009tw}, and MACRO \citep{MACRO_Ambrosio:2000yx}.  

This paper is structured as follows: Sec.~\ref{sec2} describes the detector.  
The data sample and cut parameters are discussed in Sec.~\ref{sec2andahalf}, along with the simulation.  
In Sec.~\ref{sec3} the detector performance is characterized for source searches.  
Sec.~\ref{sec4} describes the unbinned maximum likelihood search method, and in Sec.~\ref{sec5} the point-source and stacking searches are discussed.
After discussing the systematic errors in Sec.~\ref{sec6}, the results are presented in Sec.~\ref{sec7}.  
Sec.~\ref{sec8} discusses the impact of our results on various possible neutrino emission models, and Sec.~\ref{sec9} offers some conclusions.

\section{Detector and Data Sample}
\label{sec2}

The IceCube Neutrino Observatory will be composed of a deep array of 86 strings holding 5,160 digital optical modules (DOMs) deployed between 1.45 and 2.45 km below the surface of the South Pole ice.
The strings are typically separated by about 125~m with DOMs separated vertically by about 17~m along each string.  
IceCube construction started with the first string installed in the 2005--6 austral summer \citep{IceCubeFirstYearPerformance_Achterberg:2006md} and will be completed by the end of 2010.  
Six of the strings in the final detector will use high quantum efficiency DOMs and a spacing of about 70~m horizontally and 7~m vertically.  
Two more strings will have standard IceCube DOMs and 7~m vertical spacing but an even smaller horizontal spacing of 42~m.  
These eight strings along with seven neighboring standard strings make up DeepCore, designed to enhance the physics performance of IceCube below 1~TeV.  
The observatory also includes a surface array, IceTop, for extensive air shower measurements on the composition and spectrum of CRs.  

Each DOM consists of a 25 cm diameter Hamamatsu PMT \citep{Abbasi:2010vc}, electronics for waveform digitization \citep{IceCubeDAQ_:2008ym}, and a spherical, pressure-resistant glass housing.  
A single Cherenkov photon arriving at a DOM and producing a photoelectron is defined as a hit.  
A trigger threshold equivalent to 0.25 of the average single photoelectron signal is applied to the analog output of the PMT.  
The waveform of the PMT total charge is digitized and sent to the surface if hits are in coincidence with at least one other hit in the nearest or next-to-nearest neighboring DOMs within $\pm 1000$~ns.  
Hits that satisfy this condition are called local coincidence hits.  
The waveforms can contain multiple hits.
The total number of photoelectrons and their arrival times are extracted with an iterative Bayesian-based unfolding algorithm. This algorithm uses the template shape representing an average hit.

\begin{figure} [ht!]
\epsscale{0.6}
\plotone{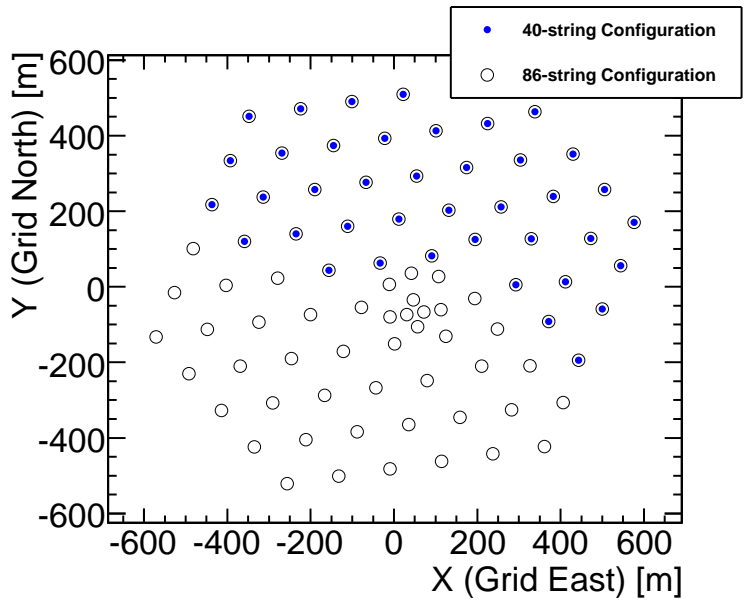}
\caption{\label{fig:layout} Overhead view of the 40-string configuration, along with additional strings that will make up the complete IceCube detector. }
\end{figure}

Forty strings of IceCube were in operation from 2008 April 5 to 2009 May 20.  
The layout of these strings in relation to the final 86-string IceCube configuration is shown in Fig.~\ref{fig:layout}.  
Over the entire period the detector ran with an uptime of 92\%, yielding 375.5 days of total exposure.  
Deadtime is mainly due to test runs during and after the construction season dedicated to calibrating the additional strings and upgrading data acquisition systems.  

IceCube uses a simple multiplicity trigger, requiring local coincidence hits in eight DOMs within $5\,\mu$s.  
Once the trigger condition is met, local coincidence hits within a readout window $\pm 10\,\mu$s are recorded, and overlapping readout windows are merged together.  
IceCube triggers primarily on down-going muons at a rate of about 950~Hz in this (40-string) configuration.  
Variations in the trigger rate by about $\pm10\%$ are due to seasonal changes affecting development of CR showers and muon production in the atmosphere, with higher rates during the austral summer \citep{Tilav:2010hj}.  

\section{Data and Simulation}
\label{sec2andahalf}
\subsection{Data Sample}
\label{sub:data}

Traditional astrophysical neutrino point source searches have used the Earth to block all upward traveling (up-going) particles except muons induced by neutrinos, as in \citet{Abbasi:2009iv}.  
There remains a background of up-going muons from neutrinos, which are created in CR air showers and can penetrate the entire Earth.
These atmospheric neutrinos have a softer energy spectrum 
than many expectations for astrophysical neutrinos.  
The measurement of the atmospheric neutrino spectrum for the 40-string detector is discussed in \citet{Abbasi:2010ie}.
A large number of muons produced in CR showers in the atmosphere and moving downward through the detector (down-going) are initially mis-reconstructed as up-going.
These mask the neutrino events until quality selections are made, leaving only a small residual of mis-reconstructed events.

The down-going region is dominated by atmospheric muons that also have a softer spectrum compared to muons induced by astrophysical neutrinos.  
At present, this large background reduces the IceCube sensitivity to neutrino sources in the southern sky in the sub-PeV energy region.  
While veto techniques are in development which will enable larger detector configurations to isolate neutrino-induced events starting within the detector, point source searches can meanwhile be extended to the down-going region if the softer-spectrum atmospheric muon background is reduced by an energy selection.  
This was done for the first time using the previous 22-string configuration of IceCube \citep{Abbasi:2009cv}, extending IceCube's field of view to $-50^{\circ}$ declination.  
In this paper we extend the field of view to $-85^{\circ}$ declination (the exclusion between $-85^{\circ}$ and $-90^{\circ}$ is due to the use of scrambled data for background estimation in the analysis, described in Sec.\,\ref{sec4}).
Down-going  muons can also be created in showers caused by gamma rays, which point back to their source like neutrinos.  
The possibility for IceCube to detect PeV gamma ray sources in the southern sky is discussed in \citet{Halzen:2009us}, which concludes that a realistic source could be detected using muons in the ice only after 10 years of observing.
Gamma ray sources will not be considered further here.   

Two processing levels are used to reduce the approximately $3.3\times10^{10}$ triggered events down to a suitable sample for analysis (see Table~\ref{tab:nevents}).  
Random noise at the level of about 500 Hz per DOM is mainly due to radioactive decays in the materials in the DOMs. 
The contribution to triggered events by this random noise is highly suppressed by the local coincident hit requirement.
To further reduce the contribution from noise, only hits within a $6~\mu$s time window are used for the reconstructions.
This time window is defined as the window that contains the most hits during the event.  
About 5\% of down-going muons which trigger the detector are initially mis-reconstructed as up-going by the first stages of event processing.
A persistent background that grows with the size of the detector is CR muons (or bundles of muons) from different showers which arrive in coincidence. 
At trigger level they make up about 13\% of the events.
These coincident muon bundles can mimic the hit pattern of good up-going events, confusing a single-muon fit.  

A likelihood-based muon track reconstruction is first performed at the South Pole (L1 filter).  
The likelihood function \citep{Ahrens:2003fg} parameterizes the probability of observing the geometry and timing of the hits in terms of a muon track's position, zenith angle, and azimuth angle.  
This likelihood is maximized, yielding the best-fit direction and position for the muon track.  
Initial fits are performed using a single photoelectron (SPE) likelihood that uses the time of the leading edge of the first photon arriving in each DOM.  
These reconstructions yield robust results used for the first level of background rejection.  
All events that are reconstructed as up-going are kept, while events in the down-going region must pass an energy cut that tightens with decreasing zenith angle. 
Events pass this L1 filter at an average rate of about $22\,\mathrm{Hz}$ and are buffered before transmission via a communications satellite using the South Pole Archival and Data Exchange (SPADE) system.  


\begin{deluxetable}{c  c}
\tablewidth{0pt}
\tablecaption{\label{tab:nevents}  Number of events at each processing level for the 375.5 d of livetime.}
\startdata
\hline\hline
Triggered Events &  $3.3 \times 10^{10}$ \\ 
\hline
 L1 Filtered Events &  $8.0 \times 10^{8}$ \\
 \hline
 Events in Final Sample & 36,900 \\
\enddata
\end{deluxetable}

\begin{figure} [ht!]
\epsscale{0.5}
\plotone{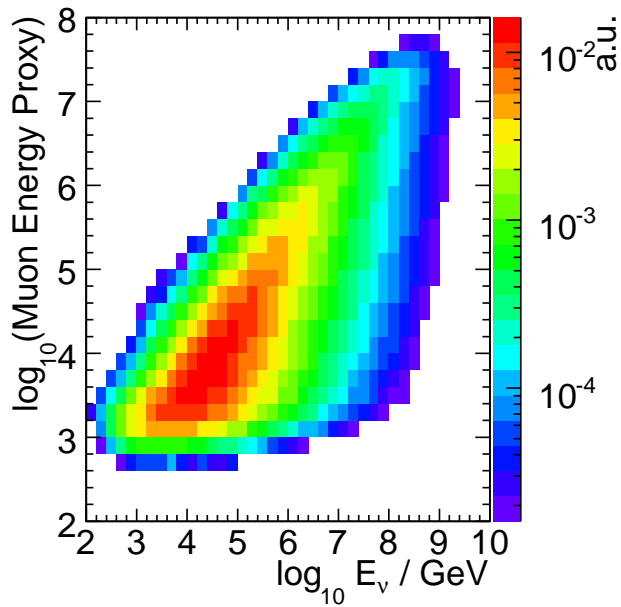}
\caption{\label{fig:muevsenergy} Distribution of the muon energy proxy (energy loss observed in the detector) versus the true neutrino energy for an arbitrary normalization $E^{-2}$ flux.}
\end{figure}

The processing done in the North includes a broader base of reconstructions compared to what is done at the South Pole.  
Rather than just the simple SPE fit, the multiple photoelectron (MPE) fit uses the number of observed photons to describe the expected arrival time of the first photon.  
This first photon is scattered less than an average photon when many arrive at the same DOM.  
The MPE likelihood description uses more available information than SPE and improves the tracking resolution as energy increases, and this reconstruction is used for the final analysis.  
The offline processing also provides parameters useful for background rejection, reconstructs the muon energy, and estimates the angular resolution on an event-by-event basis.  
Reducing the filtered events to the final sample of this analysis requires cutting on the following parameters:

\begin{itemize}
\item{
{\bf Reduced log-likelihood:}  The log-likelihood from the muon track fit divided by the number of degrees of freedom, given by the number of DOMs with hits minus five, the number of free parameters used to describe the muon. 
This parameter performed poorly on low energy signal events.  
It was found that low energy efficiency could be increased by instead dividing the likelihood by number of DOMs with photon hits minus 2.5.  
Both the standard and modified parameters were used, requiring events to pass one selection or the other.  
This kept the efficiency higher for a broader energy range.
}
\item{
{\bf Angular uncertainty, $\sigma$:}  An estimate of the uncertainty in the muon track direction.  
The directional likelihood space around the best track solution is sampled and fit to a paraboloid.  
The contour of the paraboloid traces an error ellipse indicating how well the muon direction is localized \citep{Paraboloid_Neunhoffer:2004ha}.
The RMS of the major and minor axes of the error ellipse is used to define a circular error.  
This parameter is effective both for removing mis-reconstructed events and as an event-by-event angular uncertainty estimator.
}
\item{
{\bf Muon energy proxy:} The average photon density along the muon track, used as a proxy for the muon energy.  
It is calculated accounting for the distance to DOMs, their angular acceptance, and average scattering and absorption properties of photons in the ice.  
The energy loss of a muon moving through the detector scales with the muon energy above about 1~TeV when stochastic energy losses due to bremsstrahlung, pair production, and photonuclear interactions dominate with respect to ionization losses.  
The energy resolution obtained is of the order of 0.3 in the log$_{10}$ of the muon energy (at closest approach to the average hit location) for energies between about 10~TeV and 100~PeV. 
The energy of the parent neutrino can be inferred from the reconstructed muon energy loss in the detector using simulation.
Since the interaction vertex is often an unknown distance from the detector, the muon in the detection volume has already lost an unknown amount of energy.  
Fig.~\ref{fig:muevsenergy} shows the distribution of this energy parameter versus the true neutrino energy for an $E^{-2}$ spectrum.    
Despite the uncertainty on the neutrino energy, for a statistical sample of events this energy estimator is a powerful analysis tool because of the wide range over which energies are measured. 
}
\item{
{\bf Zenith-weighted likelihood ratio:} The likelihood ratio between an unbiased muon fit and a fit with an event weight according to the known down-going muon zenith distribution as a Bayesian prior.
Applied to up-going tracks, a high likelihood ratio establishes strong evidence that the event is actually up-going and not a mis-reconstructed down-going event.  
}
\item{
{\bf NDir:} The number of DOMs with direct photons, defined as arriving within -15 ns  to +75 ns of the expectation from an unscattered photon emitted from the reconstructed muon track at the Cherenkov angle.  Scattering of photons in the ice causes a loss of directional information and will delay them with respect to the unscattered expectation.
}
\item{
{\bf LDir:} The maximum length between direct photons, projected along the best muon track solution.  
}
\item{
{\bf Zenith directions of split events:} The zenith angles resulting from splitting of an event into two parts and reconstructing each part separately. This is done in 2 ways: temporally, by using the mean photon arrival time as the split criterion; and geometrically, by using the plane both perpendicular to the track and containing the average hit location as the split criterion.  
This technique is effective against coincident muon bundles mis-reconstructed as single up-going tracks if both sub-events are required to be up-going.
}
\end{itemize}

In the up-going region, all parameters are used.  
The zenith-weighted likelihood ratio and event splitting are specifically designed to remove down-going atmospheric muon backgrounds that have been mis-reconstructed as up-going while the other parameters focus on overall track quality.  

In the down-going region, without a veto or earth filter, muons from CR showers overwhelm neutrino-induced muons, except possibly at high energies if the neutrino source spectra are harder than the CR spectrum.  
The aim of the analysis in this region is therefore to select high energy, well-reconstructed events.
We use the first three parameters in the list above as cut variables, requiring a higher track quality than in the up-going range. 
Energy cuts were introduced in the down-going region to reduce the number of events to a suitable size, cutting to achieve a constant number of events per solid angle (which also simplifies the background estimation in the analysis).  
This technique keeps the high energy events which are most important for discovery.  
Cuts were optimized for the best sensitivity using a simulated signal of muon neutrinos with $E^{-2}$ spectrum.
We checked that the same cuts resulted in a nearly optimal sensitivity for a softer $E^{-3}$ spectrum in the up-going region where low energy sensitivity is possible and for a harder $E^{-1.5}$ spectra in the down-going region.  

\begin{figure} [ht!]
\epsscale{1.0}
\plotone{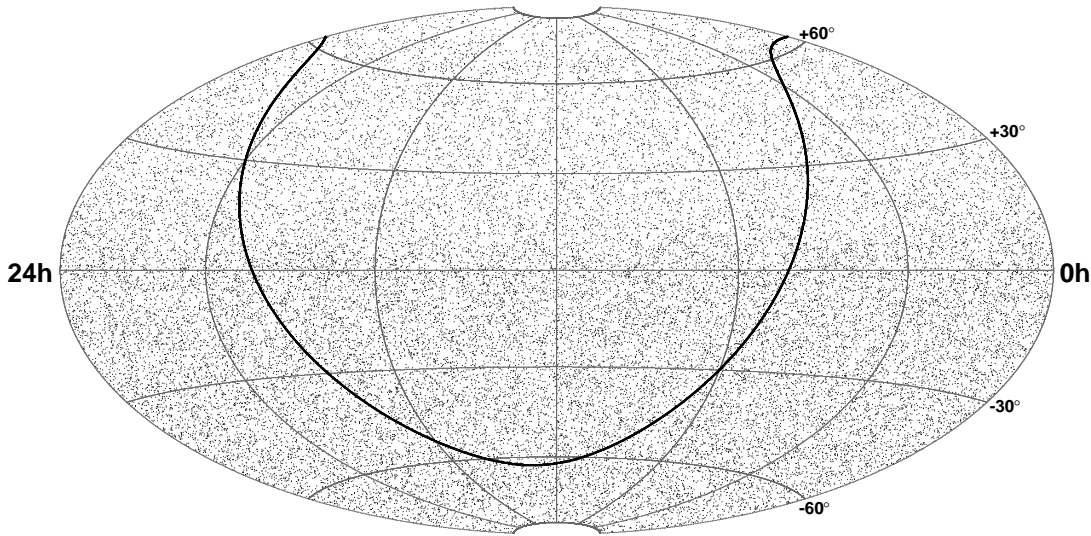}
\caption{\label{fig:eventmap}Equatorial skymap (J2000) of the 36,900 events in the final sample.  The galactic plane is shown as the solid black curve.  The northern sky (positive declinations) is dominated by up-going atmospheric neutrino-induced muons, and the southern sky (negative declinations) is dominated by muons produced in cosmic ray showers in the atmosphere above the South Pole.}
\end{figure}

Of the 36,900 events passing all selection criteria, 14,121 are up-going events from the northern sky, mostly muons induced by atmospheric neutrinos.  
Simulations of CR air showers with CORSIKA show a 2.4$\pm$0.8\% contamination due to mis-reconstructed down-going atmospheric muons.  
The other 22,779 are down-going events from the southern sky, mostly high energy atmospheric muons.  
An equatorial sky map of these events is given in Fig.~\ref{fig:eventmap}.

\subsection{Data and Simulation Comparison}
\label{sub:simu}

Simulation of neutrinos is used for determining event selection and calculating upper limits.  
The simulation of neutrinos is based on ANIS \citep{Gazizov:2004va}.  
Deep inelastic neutrino-nucleon cross sections use CTEQ5 parton distribution functions \citep{Lai:1999wy}.  
Neutrino simulation can be weighted for different fluxes, accounting for the probability of each event to occur.  
In this way, the same simulation sample can be used to represent atmospheric neutrino models such as Bartol \citep{Barr:2004br} and Honda \citep{Honda:2006qj} neutrino fluxes from pion and kaon decays (conventional flux).  
Neutrinos from charmed meson decays (prompt flux) have been simulated according to a variety of models \citep{Martin:2003us,Enberg:2008te,Naumov}.  
Seasonal variations in atmospheric neutrino rates are expected to be a maximum of $\pm4\%$ for neutrinos originating near the polar regions.  
Near the equator, atmospheric variations are much smaller and the variation in the number of events is expected to be less than $\pm 0.5\%$ \citep{Ackermann:2005icrc}.   

Atmospheric muon background is simulated mostly to guide and verify the event selection.  
Muons from CR air showers were simulated with CORSIKA \citep{Heck:1998vt} with the SIBYLL hadronic interaction model \citep{Ahn:2009wx}. 
An October polar atmosphere, an average case over the year, is used for the CORSIKA simulation, ignoring the seasonal variations of $\pm 10\%$ in event rates \citep{Tilav:2010hj}.  
Muon propagation through the Earth and ice are done using MMC \citep{Chirkin:2004hz}.  
Using measurements of the scattering and absorption lengths in ice \citep{OpticalProperties_JGeophysRes_2006}, a detailed simulation propagates the photon signal to each DOM \citep{Lundberg:2007mf}.
The simulation of the DOMs includes their angular acceptance and electronics.  
Experimental and simulated data are processed and filtered in the same way.

\begin{figure} [ht!]
\epsscale{0.7}
\plotone{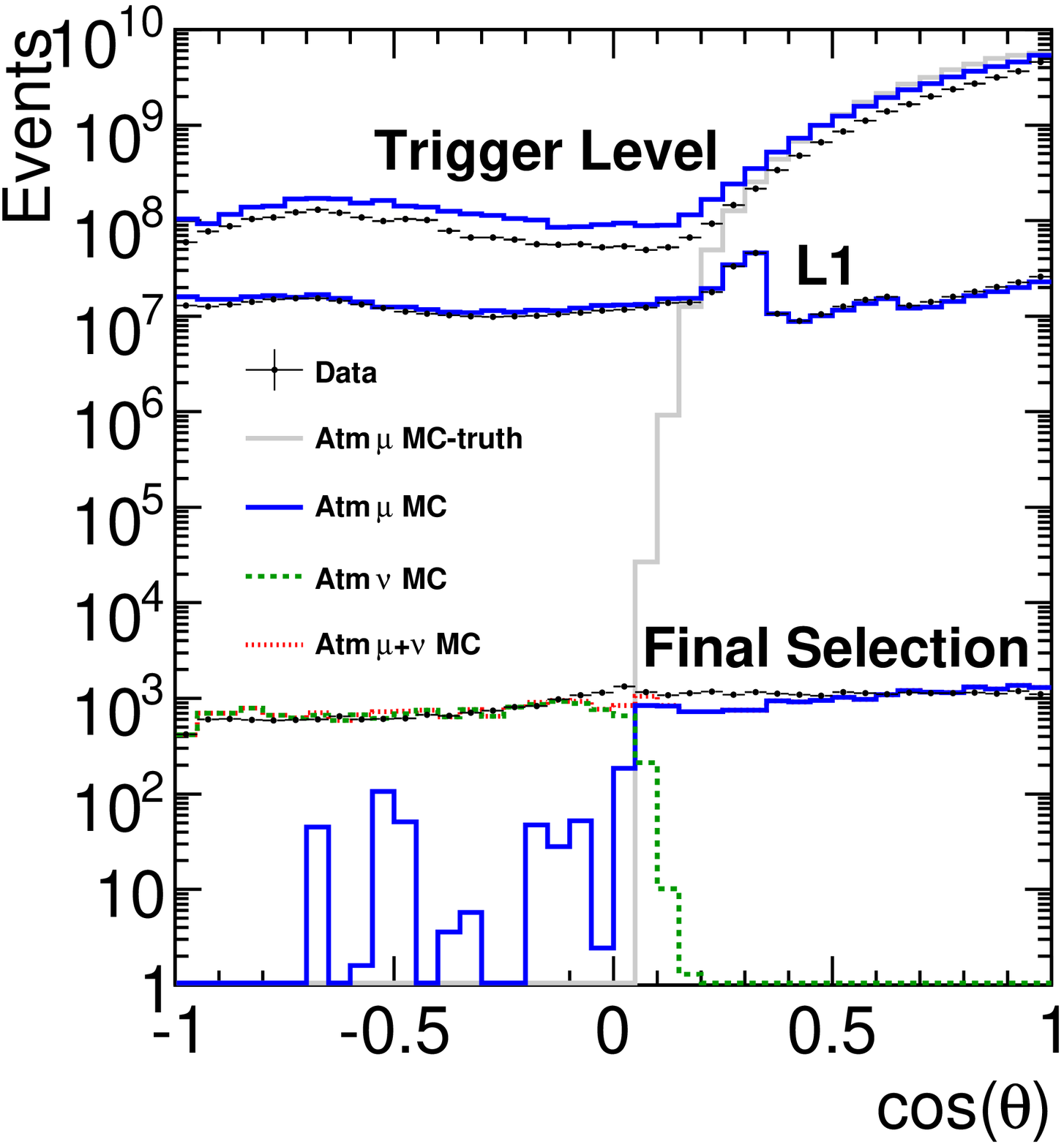}
\caption{\label{fig:zenith} Distribution of reconstructed cosine zenith at trigger level, L1, and final cut level for data and simulation of atmospheric muons \citep{Hoerandel:2002yg} and neutrinos \citep{Barr:2004br, Naumov}.  The true cosine zenith distribution of the muons at trigger level is also shown.  For the final sample, the deficit of simulation seen near the horizon is discussed in Sec.~\ref{sec6}, along with comparisons to other models.  The deficit is most likely caused by a deficiency in knowledge of the CR composition in the region around the knee. More detailed comparisons are discussed in Sec~\ref{sec6}, including comparisons to a variety of models. 
}
\end{figure}

\begin{figure} [ht!]
\centering
\begin{tabular}{cc}
\epsfig{file=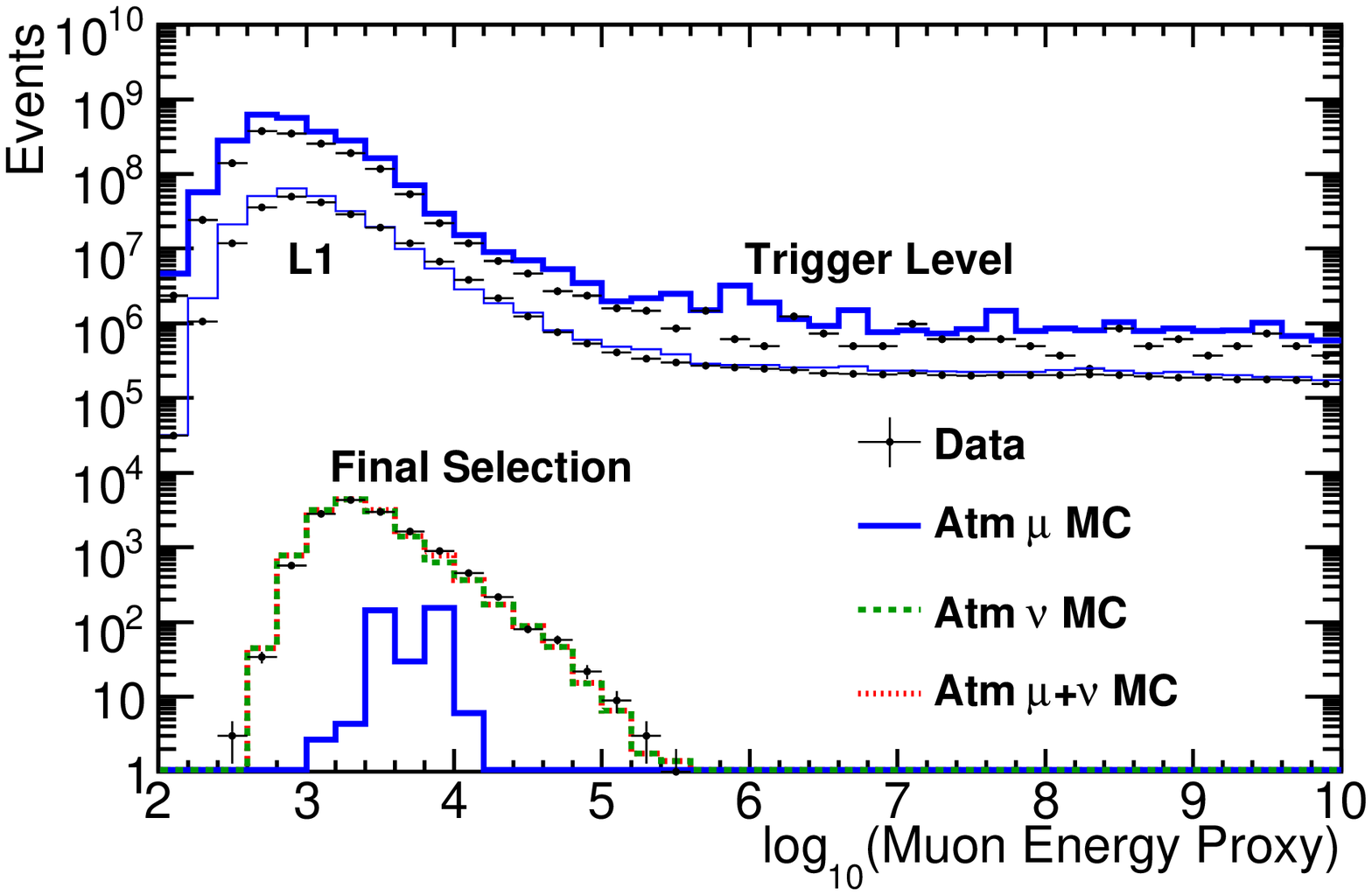,width=0.47\linewidth,clip=} &
\epsfig{file=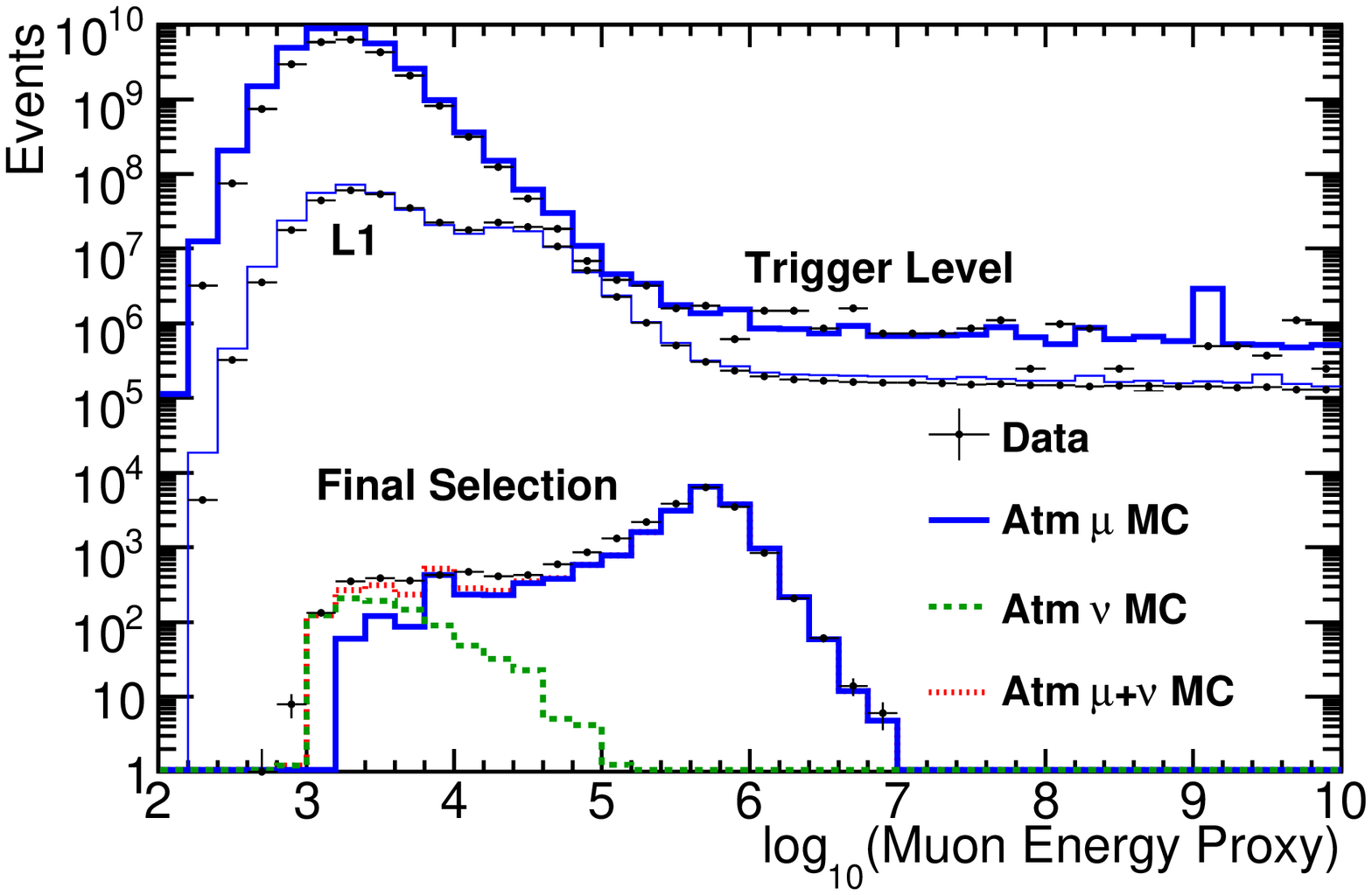,width=0.47\linewidth,clip=} \\
\epsfig{file=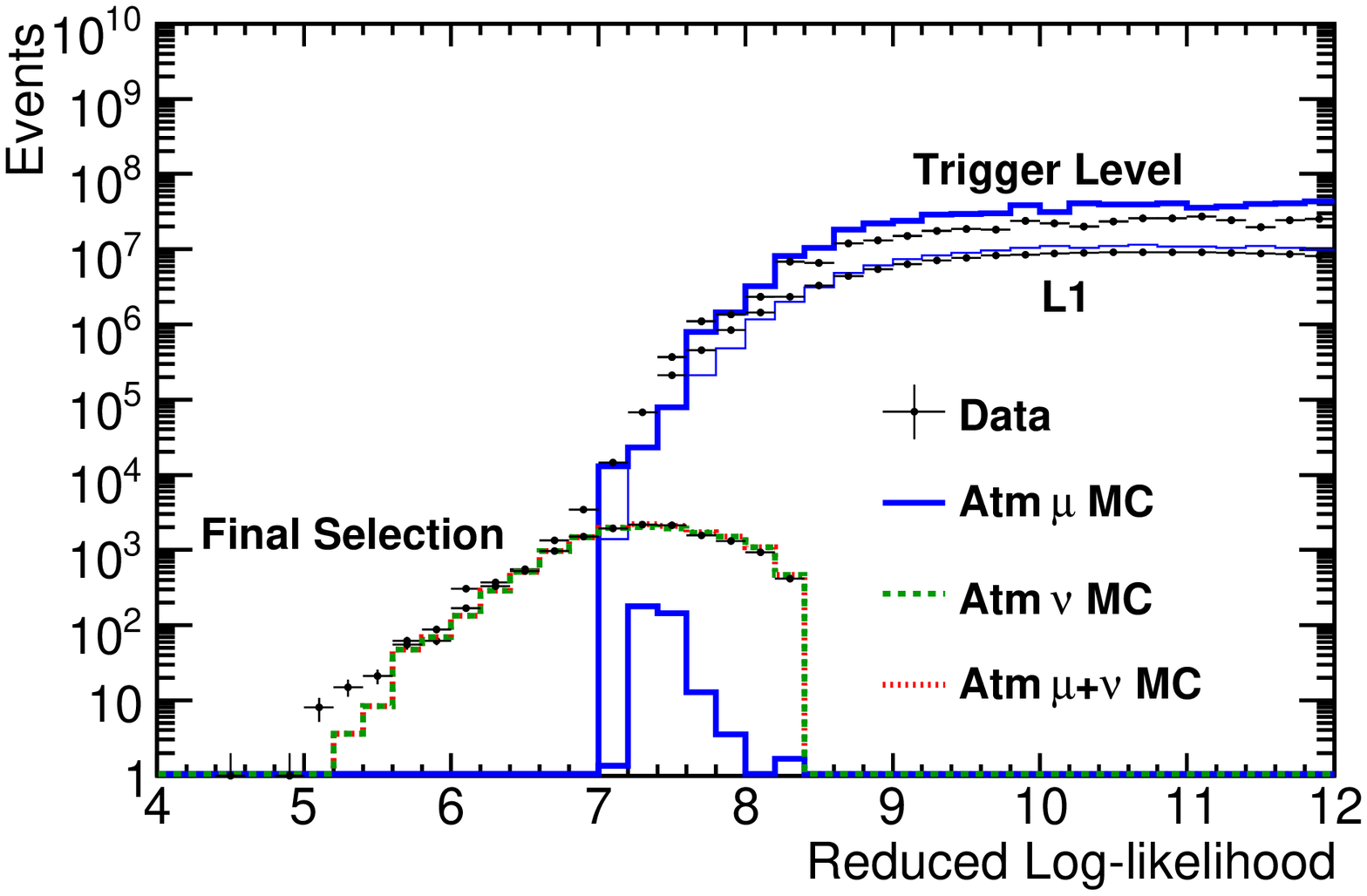,width=0.47\linewidth,clip=} & 
\epsfig{file=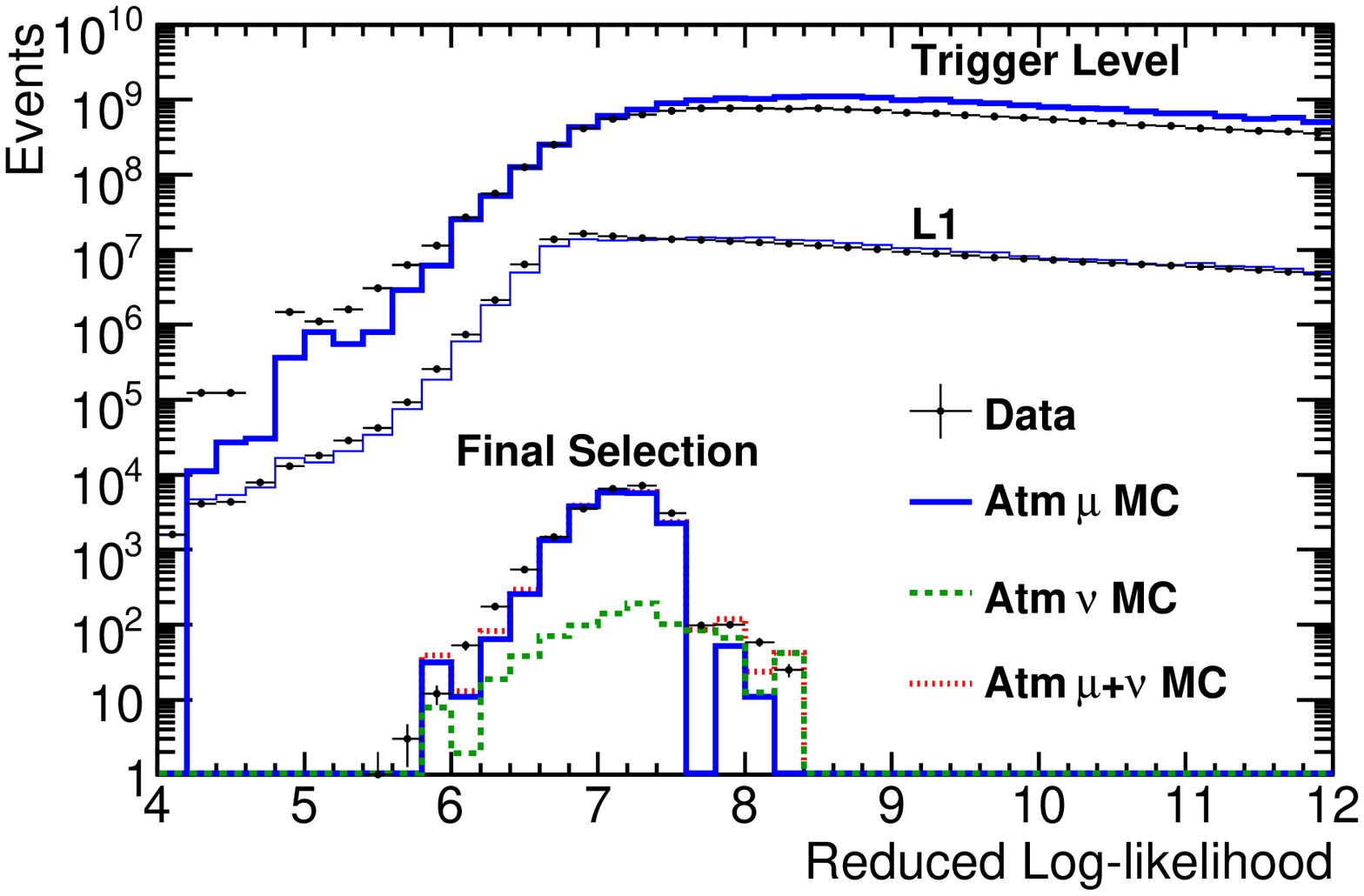,width=0.47\linewidth,clip=} \\
\epsfig{file=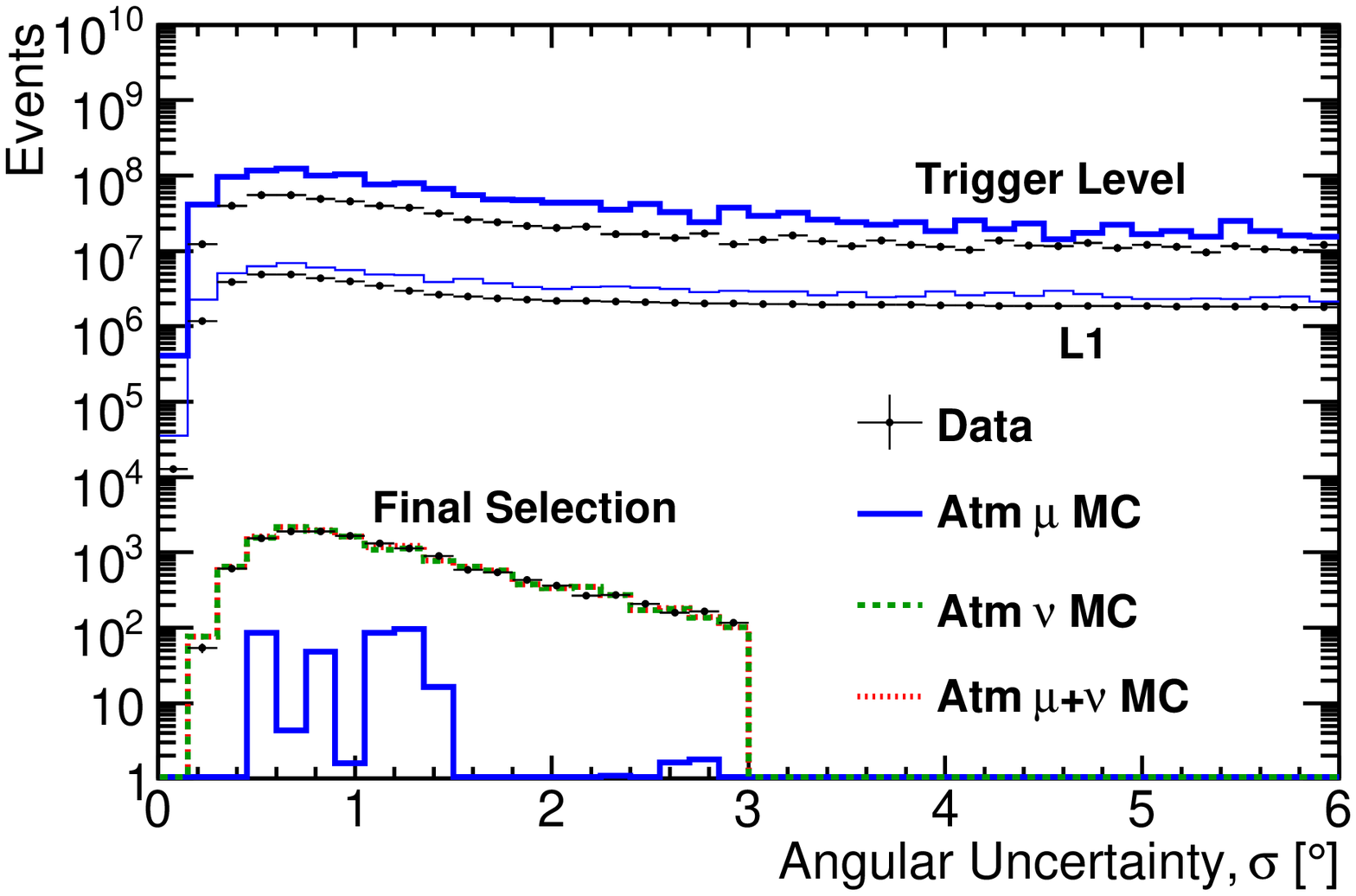,width=0.47\linewidth,clip=} &
\epsfig{file=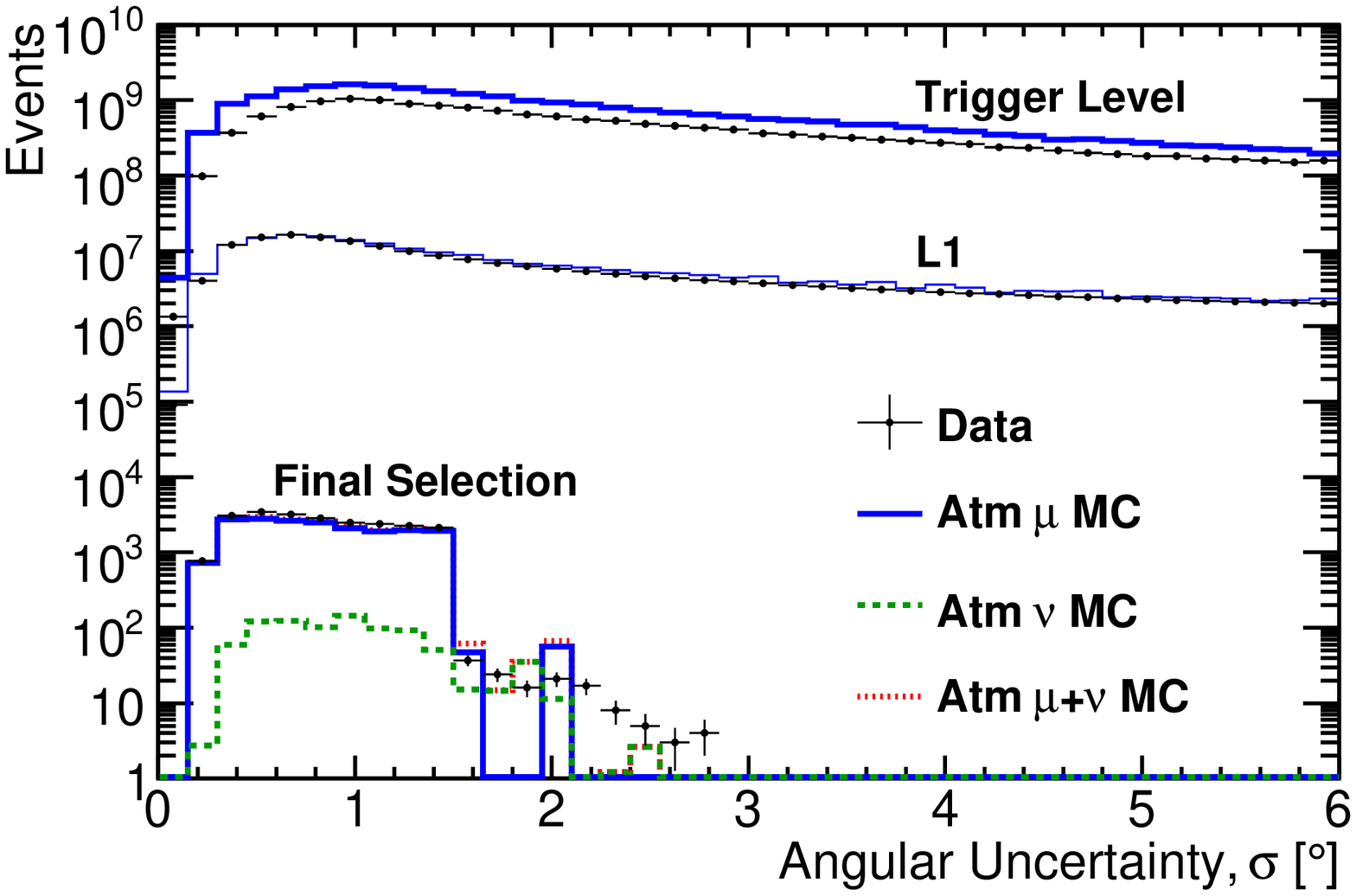,width=0.47\linewidth,clip=}
\end{tabular}
\caption{\label{fig:data_mc}  Distributions of muon energy proxy (top row), reduced log-likelihood (middle row), and angular uncertainty estimator (bottom row) for the up-going sample (left column) and the down-going sample (right column).  Each is shown at trigger level, L1, and final cut level for data and simulation of atmospheric muons and neutrinos.  
In the up-going sample (left column), all atmospheric muons are mis-reconstructed, and at final level their remaining estimated contribution is about 2.4$\pm$0.8\%.
}
\end{figure}

In Fig.~\ref{fig:zenith} we show the cosine of zenith and in Fig.~\ref{fig:data_mc} the muon energy proxy, reduced log-likelihood, and angular uncertainty estimator distributions of all events at trigger level, L1 filter level, and after final analysis cuts for data and Monte Carlo (MC).  
In these figures, the simulation uses a slightly modified version of the {\em poly-gonato} model of the galactic CR flux and composition \citep{Hoerandel:2002yg}.  
Above the galactic model cutoff at $Z \times 4\times10^{15}$~eV, a flux of pure iron is used with an $E^{-3}$ spectrum.  
This is done because currently CORSIKA cannot propagate elements in the {\em poly-gonato} model that are heavier than iron.  
Moreover, the {\em poly-gonato} model only accounts for galactic CRs and does not fully account for the average measured flux above $10^{17}$ eV, even when all nuclei are considered (see Fig.~11 in \citet{Hoerandel:2002yg}). 
These corrections then reproduce the measured CR spectrum at these energies.  
There is a 23\% difference in normalization of data and CR muon events at trigger level.  
This normalization offset largely disappears after quality cuts are made.  
Generally good agreement is achieved at later cut levels. 
Some remaining disagreements are discussed further in Sec.~\ref{sec6}, along with the impact of using different CR primary models.  

\section{Detector Performance}
\label{sec3}

\begin{figure} [ht!]
\epsscale{0.5}
\plotone{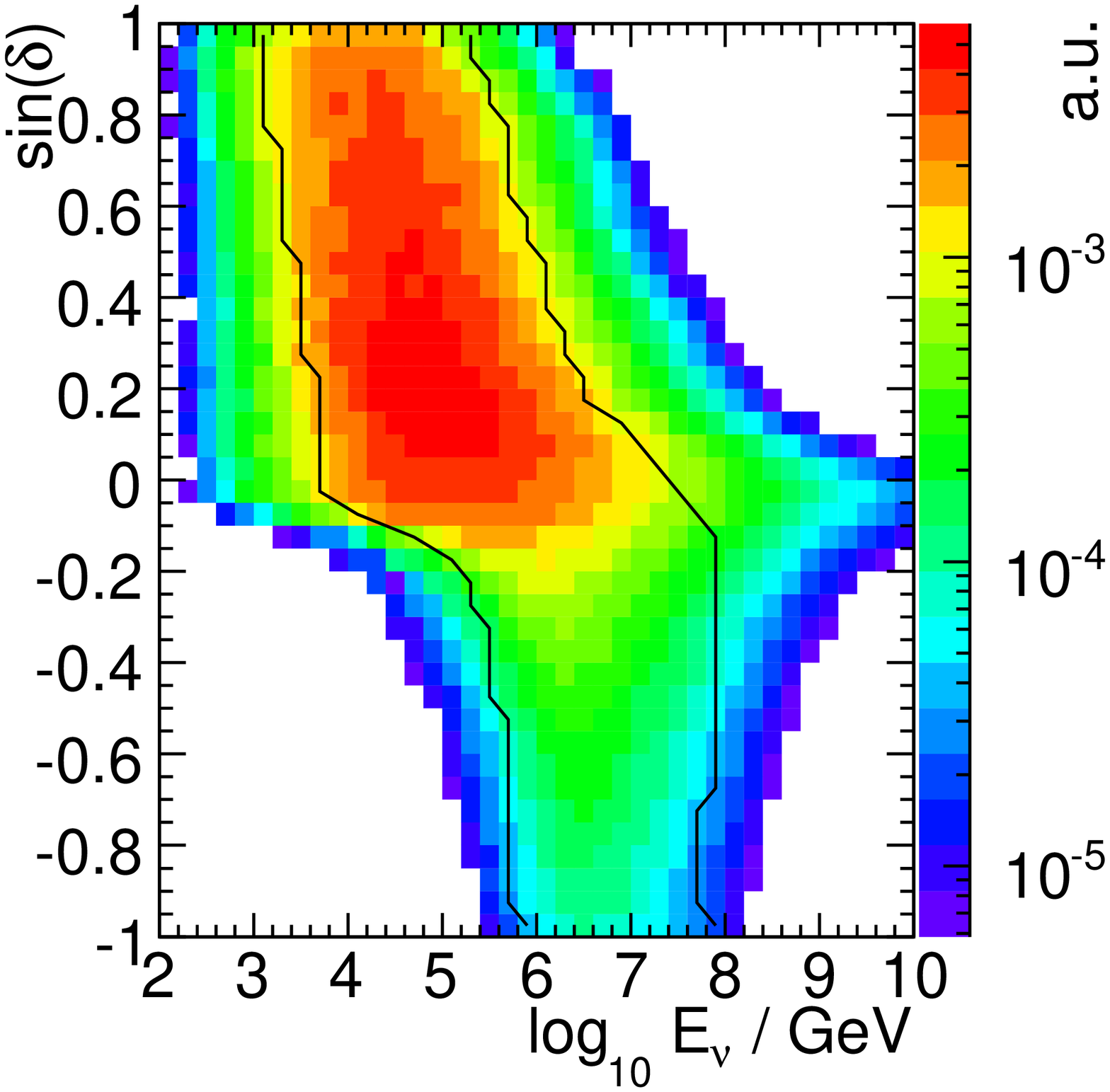}
\caption{\label{fig:evsdec} Energy distribution for an arbitrary normalization $E^{-2}$ flux of neutrinos as a function of declination for the final event selection.  The black contours indicate the 90\% central containment interval for each declination.}
\end{figure}

\begin{figure} [ht!]
\epsscale{0.6}
\begin{tabular}{cc}
\epsfig{file=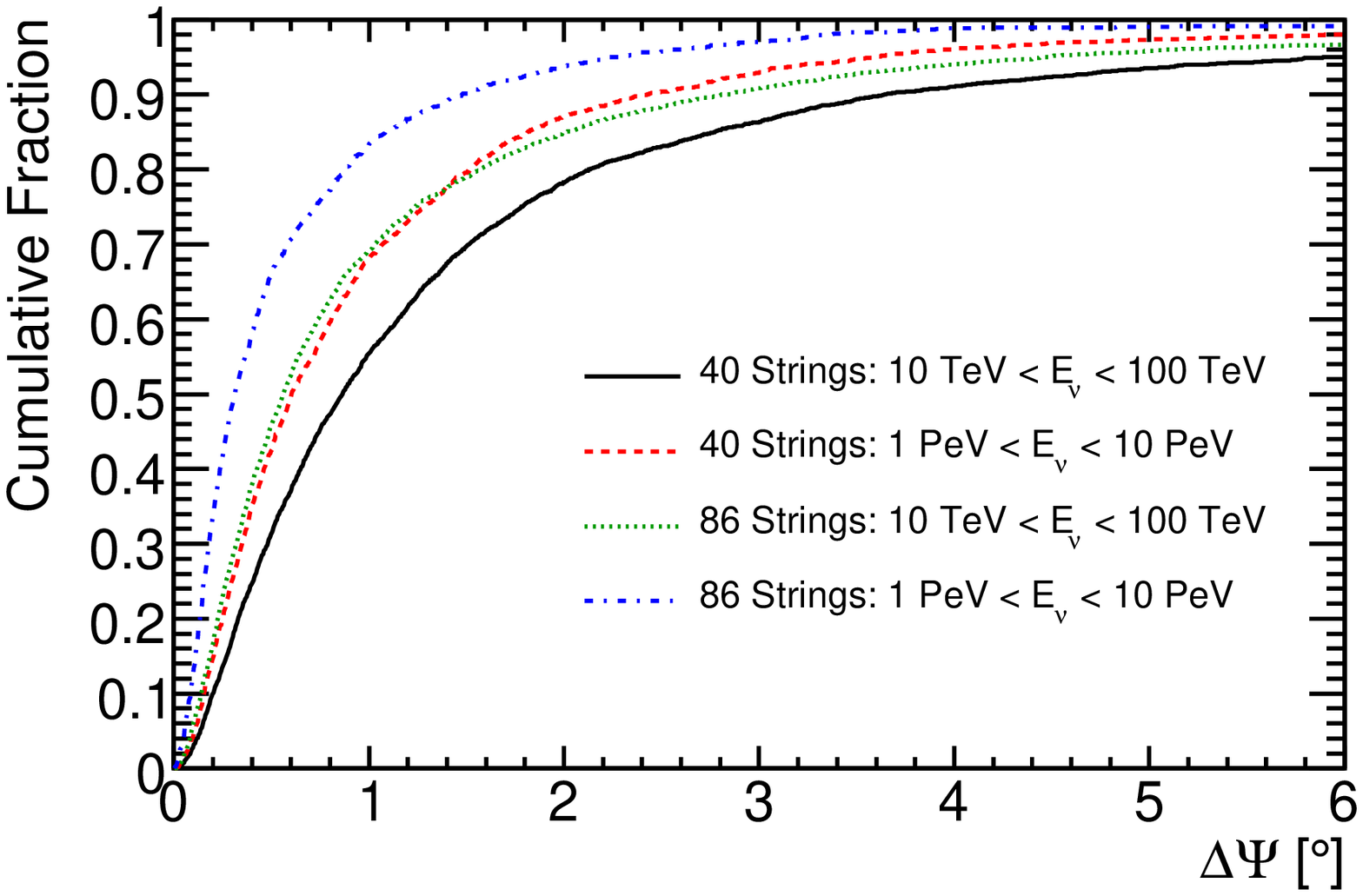,width=0.5\linewidth,clip=} &
\epsfig{file=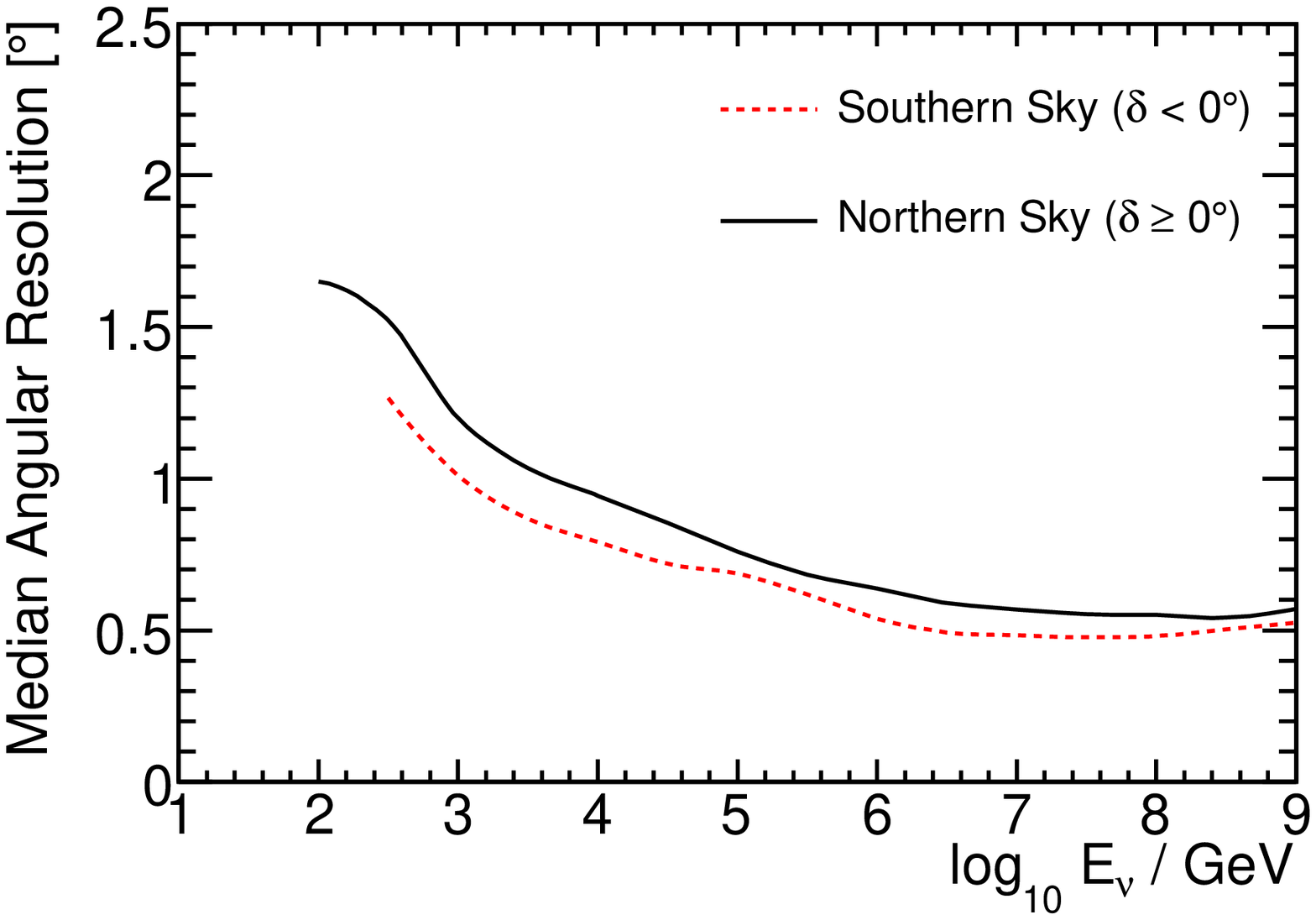,width=0.5\linewidth,clip=} \\
\end{tabular}
\caption{\label{fig:psf}  Cumulative point spread function (angle between neutrino and reconstructed muon track) for simulated $E^{-2}$ neutrino signal events at the final cut level in the up-going region (left).   Also shown is the final IceCube configuration.  The median of the PSF versus energy is shown separately for the northern and southern skies (right).  The improvement in the southern sky is because of the more restrictive quality cuts.}
\end{figure}

The performance of the detector and the analysis is characterized using the simulation described in Sec.~\ref{sub:simu}.
For an $E^{-2}$ spectrum of neutrinos, the median angular difference between the neutrino and the reconstructed direction of the muon in the northern (southern) sky is $0.8^{\circ}$ ($0.6^{\circ}$).  
Along with more severe quality selection in the southern sky, the different energy distributions in each hemisphere, shown in Fig.~\ref{fig:evsdec}, cause the difference in these two values.
This is because the reconstruction performs better at higher energy due to the larger amount of light and longer muon tracks.  
The cumulative point spread function (PSF) is shown in Fig.~\ref{fig:psf} for two energy ranges and compared with simulation of the complete IceCube detector using the same quality selection, as well as the median PSF versus energy for the two hemispheres. 

\begin{figure} [ht!]
\centering
\begin{tabular}{cc}
\epsfig{file=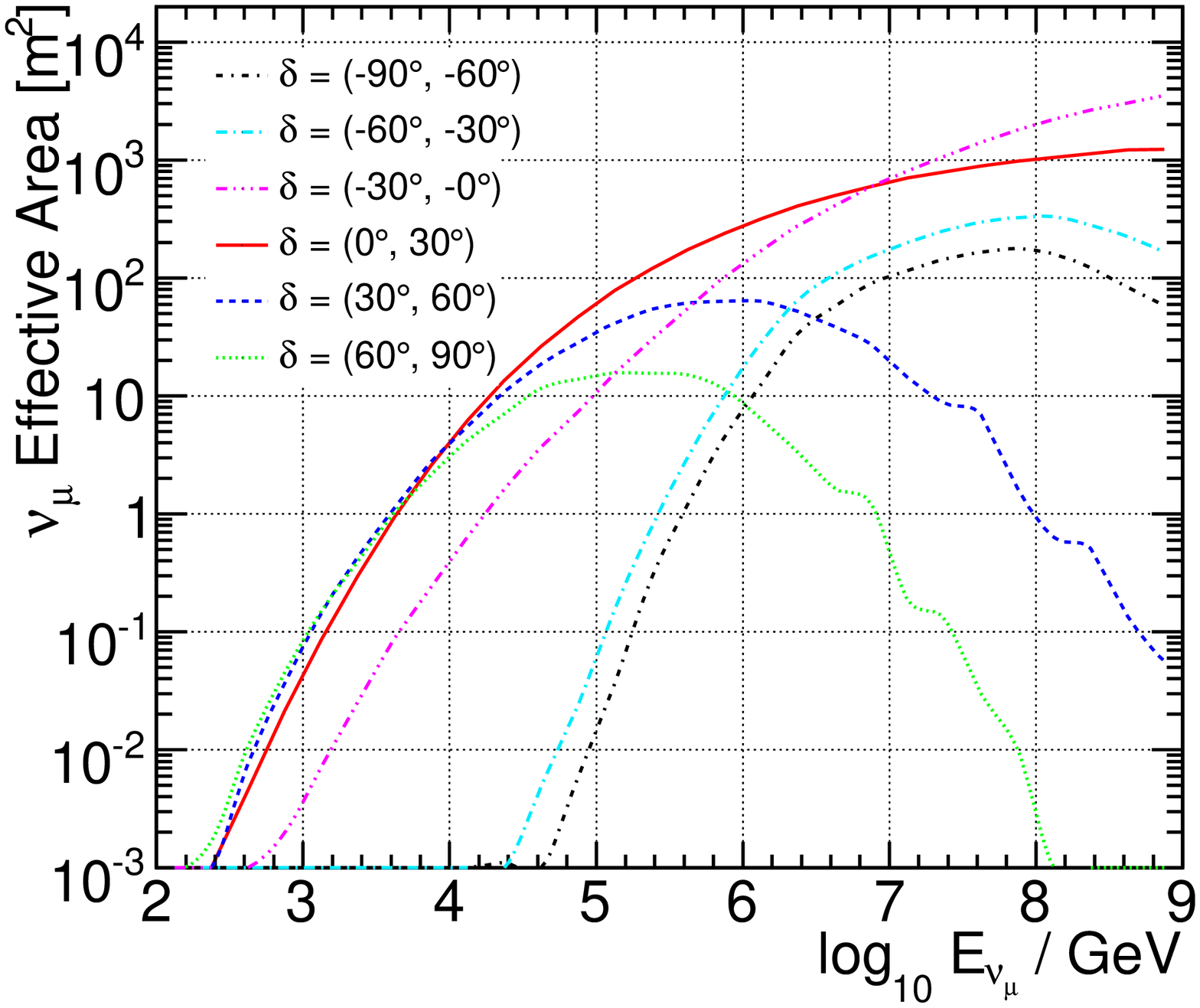,width=0.5\linewidth,clip=} &
\epsfig{file=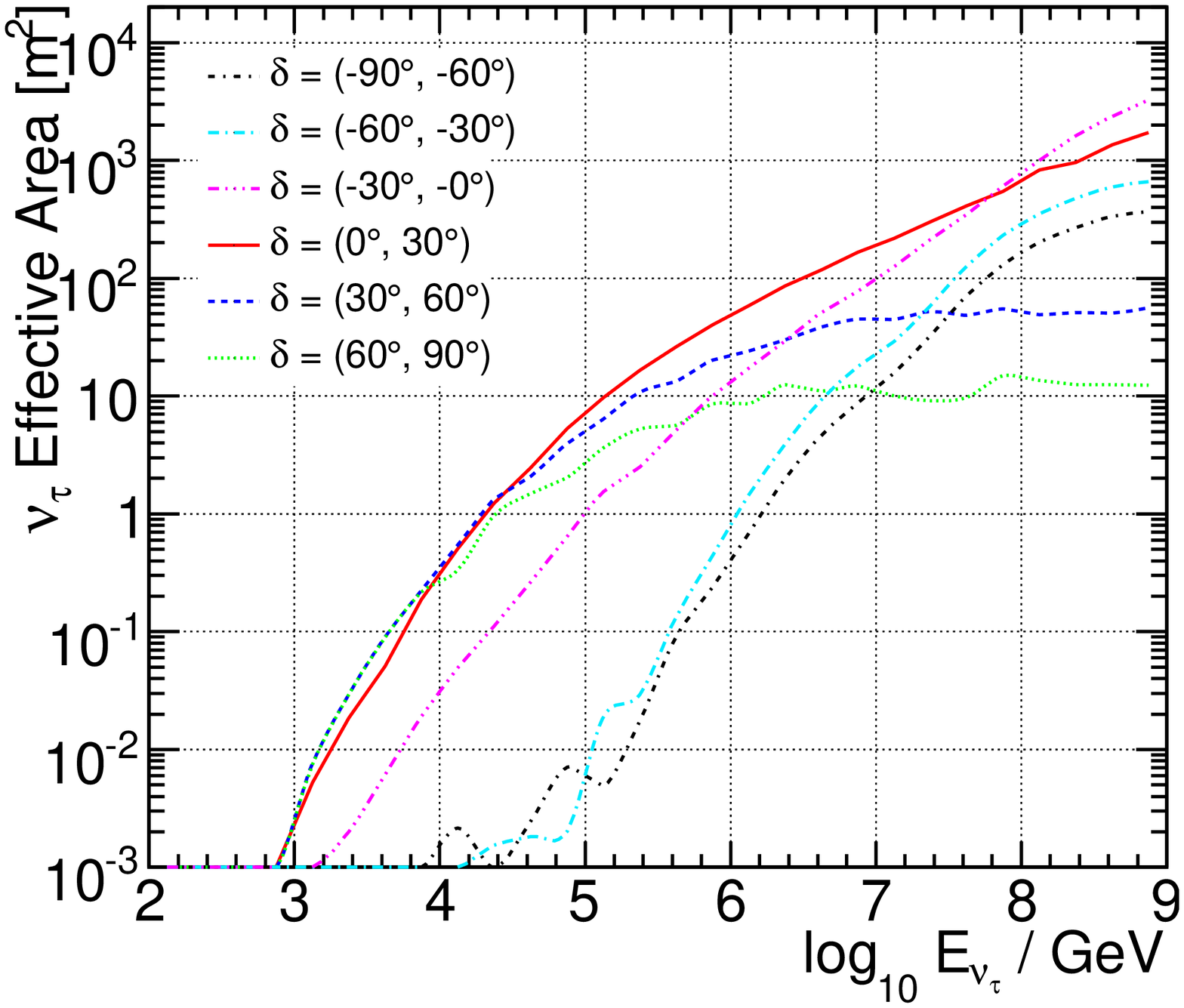,width=0.5\linewidth,clip=} \\
\end{tabular}
\caption{\label{fig:eff_area}  Solid-angle-averaged effective areas at final cut level for astrophysical neutrino fluxes in six declination bands for $\nu_{\mu} + \bar{\nu}_{\mu}$ (left) and $\nu_{\tau} + \bar{\nu}_{\tau}$ (right), assuming an equal flux of neutrinos and anti-neutrinos.  
}
\end{figure}

The neutrino effective area $A^{\rm eff}_{\nu}$
is a useful parameter to determine event rates and the performance of a detector for different analyses and fluxes. 
The expected event rate for a given differential flux $dN/dE$ is
\begin{equation}
N_{events} = \int
dE_{\nu} A^{\rm eff}_{\nu}(E_{\nu}, \delta_\nu) \frac{dN_\nu(E_{\nu}, \delta_\nu)}{dE_{\nu}} \,,
\end{equation}
and is calculable using simulation.  
The $A^{\rm eff}_{\nu}$ represents the size of an equivalent detector if it were 100\% efficient.
Figure~\ref{fig:eff_area} shows the $A^{\rm eff}_{\nu}$ for fluxes of $\nu_{\mu}+\bar{\nu}_{\mu}$ and $\nu_{\tau}+\bar{\nu}_{\tau}$, for events at final selection level.
Neutrinos arriving from the highest declinations must travel through the largest column depth and can be absorbed: this accounts for the turnover at high energies for nearly vertical up-going muon neutrinos.  
Tau neutrinos have the advantage that the tau secondary can decay back into a tau neutrino before losing much energy.  

Although tau (and electron) neutrino secondaries usually produce nearly-spherical showers rather than tracks, tau leptons will decay to muons with a 17.7\% branching ratio \citep{Amsler:2008zzb}.  
At very high energy (above about 1~PeV) a tau will travel far enough before decaying that the direction can be reconstructed well, contributing to any extraterrestrial signal in the muon channel.  
For the upper limits quoted in Sec.~\ref{sec7}, we must make an assumption on the flavor ratios at Earth, after oscillations.  
It is common to assume $\Phi_{\nu_e}:\Phi_{\nu_{\mu}}:\Phi_{\nu_{\tau}} = 1:1:1$.
This is physically motivated by neutrino production from pion decay and the subsequent muon decay, yielding $\Phi_{\nu_e}:\Phi_{\nu_{\mu}}:\Phi_{\nu_{\tau}} = 1:2:0$.  
After standard oscillations over astrophysical baselines, this gives an equal flux of each flavor at Earth \citep{Athar:2000yw}.  
Under certain astrophysical scenarios, the contribution from muon decay may be suppressed, leading to an observed flux ratio of $\Phi_{\nu_e}:\Phi_{\nu_{\mu}}:\Phi_{\nu_{\tau}} = 1:1.8:1.8$ \citep{Kashti:2005qa}, or the contribution of tau neutrinos could be enhanced by the decay of charmed mesons at very high energy \citep{Enberg:2008jm}.
For an $E^{-2}$ spectrum and equal muon and tau neutrino fluxes, the fraction of tau neutrino-induced events is about 17\% for vertically down-going, 10\% near the horizon, and 13\% for vertically up-going.
Because the contribution from tau neutrinos is relatively small, assuming only a flux of muon neutrinos can be used for convenience and to compare to other published limits.  
We have tabulated limits on both $\Phi_{\nu_{\mu}}$ and the sum $\Phi_{\nu_{\mu}}+\Phi_{\nu_{\tau}}$, assuming an equal flux of each, while in the figures we have specified that we only consider a flux of muon neutrinos.  
Limits are always reported for the flux at the surface of the Earth.   

\section{Search Method}
\label{sec4}

An unbinned maximum likelihood ratio method is used to look for a localized, statistically significant excess of events above the background.
We also use energy information to help separate possible signal from the known backgrounds.  

The method follows that of \citet{Braun:2008bg_Methods}.  The data is modeled as a two component mixture of signal and background.  
A maximum likelihood fit to the data is used to determine the relative contribution of each component.  
Given $N$ events in the data set, the probability density of the $i^{th}$ event is
\begin{equation}
\frac{n_s}{N}\mathcal{S}_i + (1 - \frac{n_s}{N})\mathcal{B}_i,
\end{equation}
where $\mathcal{S}_i$ and $\mathcal{B}_i$ are the signal and background probability density functions (PDFs), respectively. The parameter $n_s$ is the unknown contribution of signal events.

\begin{figure} [ht!]
\epsscale{0.6}
\plotone{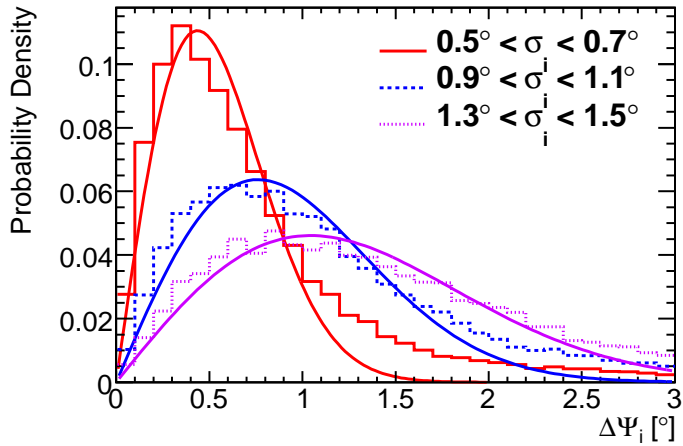}
\caption{\label{fig:parab}  Angular deviation between neutrino and reconstructed muon direction $\Delta\Psi$ for ranges in $\sigma_i$, the reconstructed angular uncertainty estimator.  Fits of these distributions to 2-dimensional Gaussians projected into $\Delta\Psi$ 
are also shown.  
The value of $\sigma_i$ is correlated to the track reconstruction error.  
A small fraction of events are not well-represented by the Gaussian distribution, but these are the least well-reconstructed events and contribute the least to signal detection.
}
\end{figure}

For an event with reconstructed direction $\vec{x}_i=(\alpha_i,\delta_i)$, we model the probability of originating from the source at $\vec{x}_s$ as a circular 2-dimensional Gaussian, 
\begin{equation}
\mathcal{N}(\vec{x}_i) = \frac{1}{2\pi \sigma_i^2}e^{-{\frac{|\vec{x}_i - \vec{x}_s|^2}{2\sigma_i^2}}},
\end{equation}
where $\sigma_i$ is the angular uncertainty reconstructed for each event individually \citep{Paraboloid_Neunhoffer:2004ha} and $|\vec{x}_i - \vec{x}_s|$ is the space angle difference between the source and reconstructed event.
%
The PSF for different ranges of $\sigma_i$ are in Fig.~\ref{fig:parab}, showing the correlation between the estimated angular uncertainty and track reconstruction error.

\begin{figure} [ht!]
\epsscale{1.0}
\plottwo{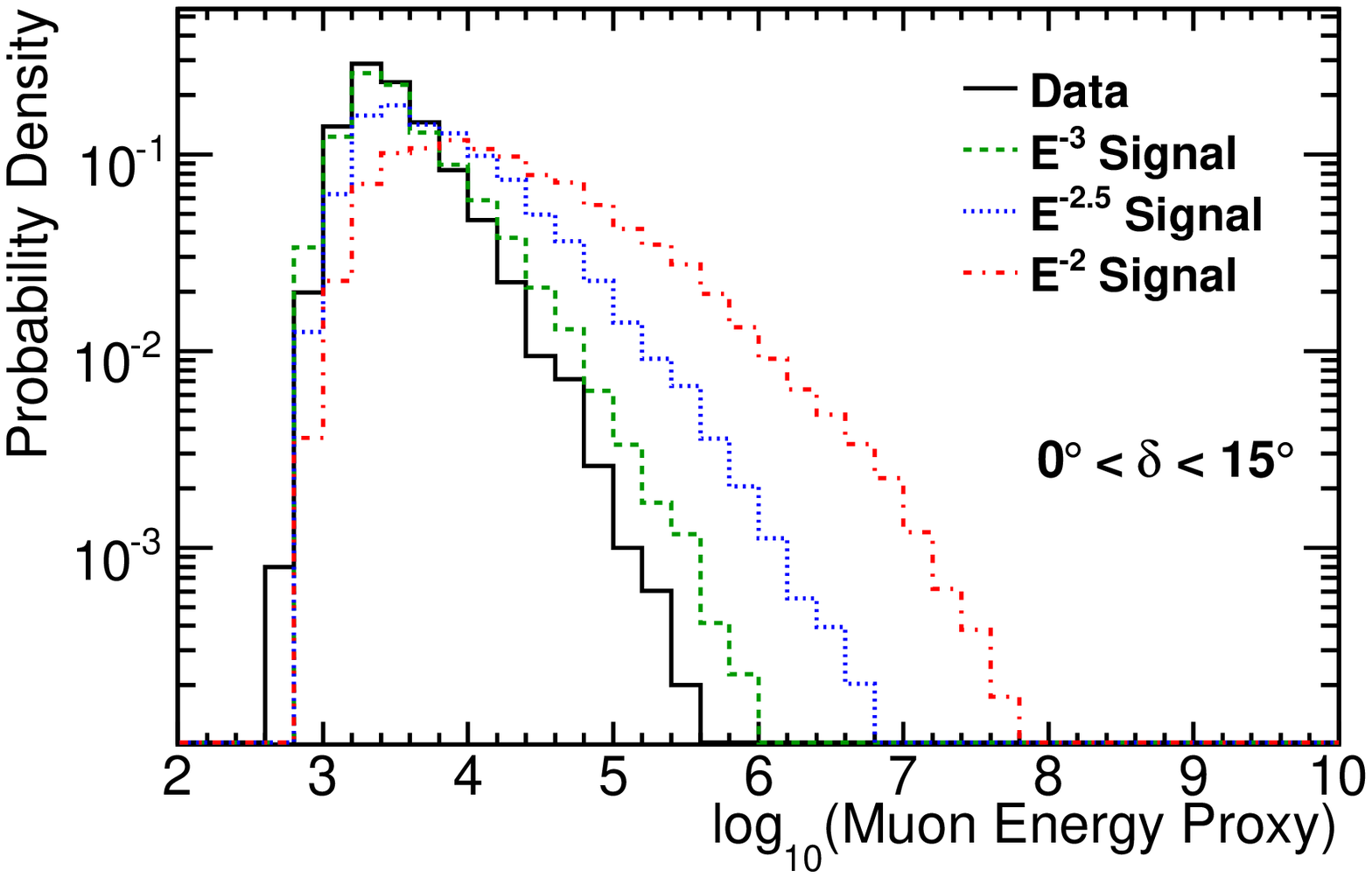}{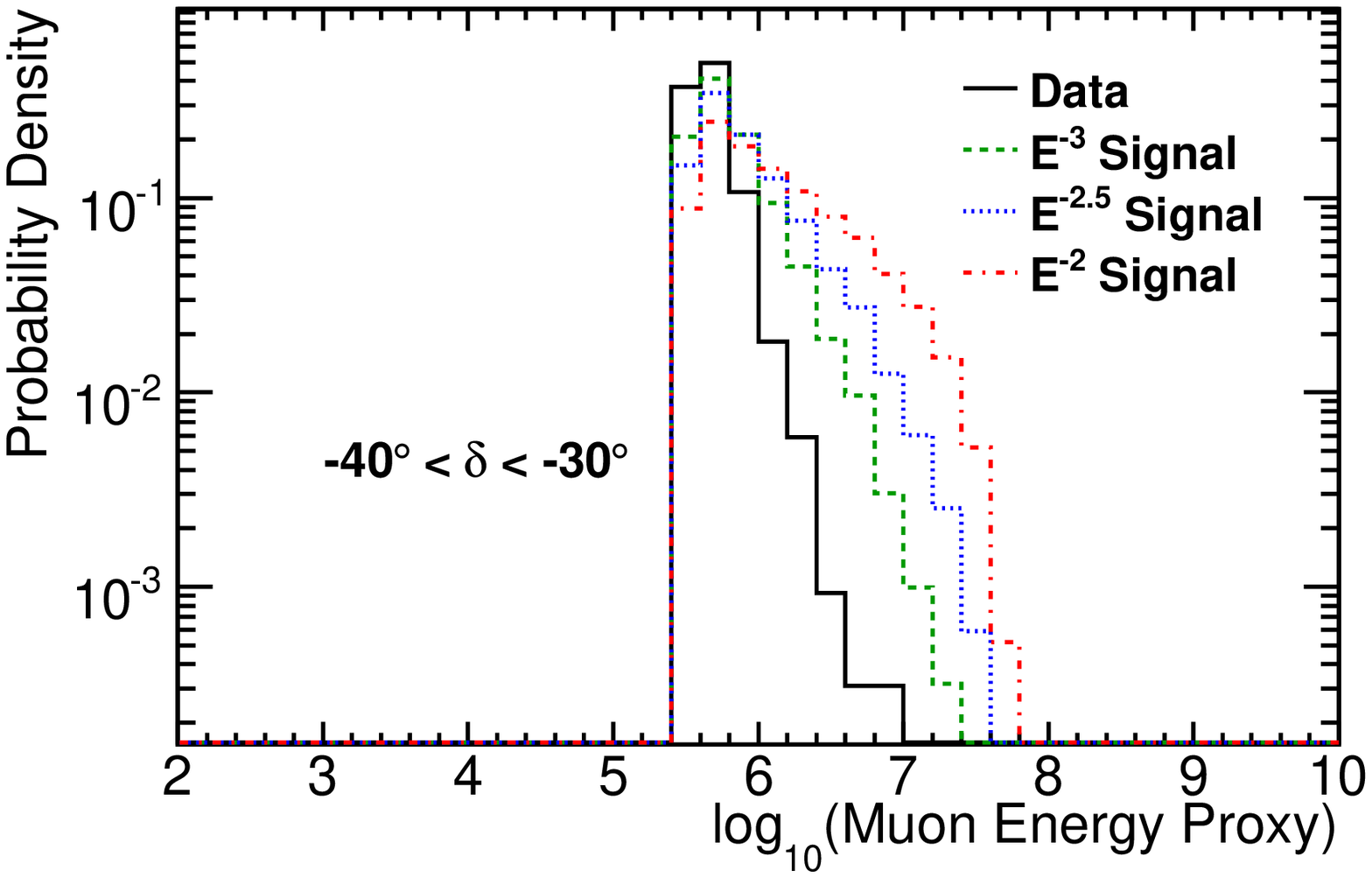}
\caption{\label{fig:eprob}  Probability densities for the muon energy proxy for data as well as simulated power-law neutrino spectra.  Two declination bands are shown: $0^{\circ} < \delta < 15^{\circ}$ (left) and $-40^{\circ} < \delta < -30^{\circ}$ (right), representing two of the declination-dependent energy PDFs used in the likelihood analysis.  There is an energy cut applied for negative declinations.
}
\end{figure}

The energy PDF $\mathcal{E}(E_i|\gamma,\delta_i)$ describes the probability of obtaining a reconstructed muon energy $E_i$ for an event produced by a source of a given neutrino energy spectrum $E^{-\gamma}$ at declination $\delta_{i}$.  
We describe the energy distribution using 22 declination bands.  
Twenty bands, spaced evenly by solid angle, cover the down-going range where the energy distributions are changing the most due to the energy cuts in the event selection, while two are needed to sufficiently describe the up-going events, with the separation at $\delta=15^{\circ}$.
We fit the source spectrum with a power law $E^{-\gamma}$; $\gamma$ is a free parameter.
The probability of obtaining a reconstructed muon energy $E_i$ for an event produced by a source with spectral index $\gamma$, for spectral indices 1.0 $<$ $\gamma$ $<$ 4.0, is determined using simulation.  
Two examples of these energy PDFs are shown in Fig.~\ref{fig:eprob}.  

The full signal PDF is given by the product of the spatial and energy PDFs: 

\begin{equation}
\mathcal{S}_i = \mathcal{N}(\vec{x}_i) \cdot \mathcal{E}(E_i|\gamma,\delta_i).
\end{equation}

The background PDF $\mathcal{B}_i$ contains the same terms, describing the angular and energy distributions of background events:
\begin{equation}
\mathcal{B}_i = \mathcal{N}_{Atm}(\vec{x}_i) \cdot \mathcal{E}(E_i|Atm,\delta_i),
\label{eq:bgpdf}
\end{equation}
where $\mathcal{N}_{Atm}(\vec{x}_i)$ is the spatial PDF of atmospheric background and $\mathcal{E}(E_i|Atm,\delta_i)$ is the probability of obtaining $E_i$ from atmospheric backgrounds (neutrinos and muons) at the declination of the event.
These PDFs are constructed using data and, for the energy term, in the same 22 declinations bands as the signal PDF.  
All non-uniformities in atmospheric background event rates caused by the detector acceptance or seasonal variation average out in the time-integrated analysis.  
Therefore $\mathcal{N}_{Atm}(\vec{x}_i)$ has a flat expectation in right ascension and is only dependent on declination.  
Because the data are used in this way for background estimation, the analysis is restricted from $-85^{\circ}$ to $85^{\circ}$ declination, so that any point source signal will still be a small contribution to the total number of events in the same declination region.   

The likelihood of the data 
is the product of all event probability densities:
\begin{equation}
\mathcal{L}(n_s, \gamma) = \prod_{i=1}^{N} \Big[\frac{n_s}{N}\mathcal{S}_i + (1 - \frac{n_s}{N})\mathcal{B}_i\Big].
\label{eq:lh}
\end{equation}
The likelihood is then maximized with respect to $n_s$ and $\gamma$, giving the best fit values $\hat{n}_s$ and $\hat{\gamma}$.  
The null hypothesis is given by $n_s=0$ ($\gamma$ has no meaning when no signal is present).  
The likelihood ratio test statistic is
\begin{equation}
TS = -2 \log\Big[\frac{\mathcal{L}(n_s=0)}{\mathcal{L}(\hat{n}_s,\hat{\gamma}_s)}\Big].
\label{eq:ts}
\end{equation}

\begin{figure} [ht!]
\epsscale{.8}
\plotone{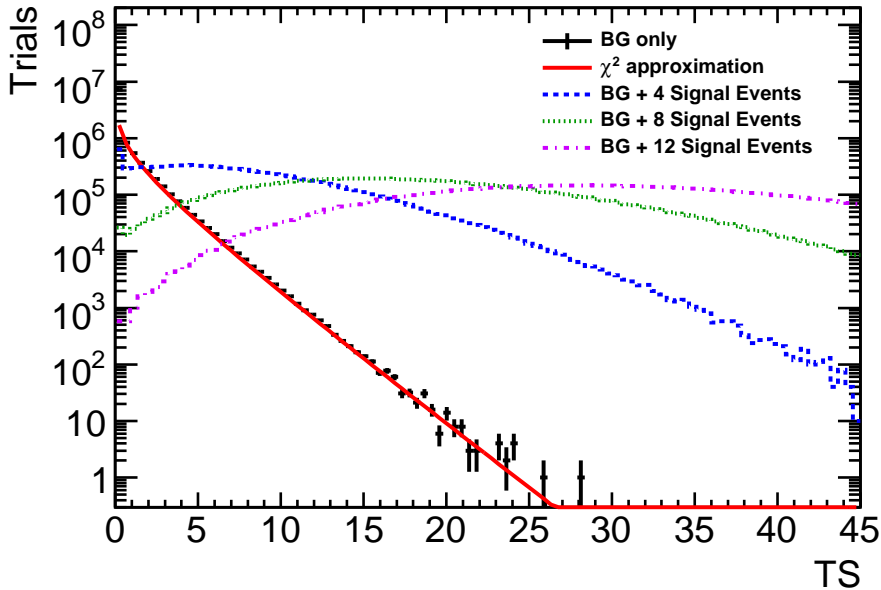}
\caption{\label{fig:loglambda}  Distribution of the test statistic $TS$ for a fixed point source at $\delta=45^{\circ}$.  Statistical errors are shown for the background-only distribution using 10 million scrambled data sets.  A $\chi^2$ distribution with 2 degrees of freedom can be used as an approximation.  Also shown are the distributions when simulated $E^{-2}$ signal events are injected.  About 12 events on average are needed for a discovery at this declination.}
\end{figure}

The significance of the analysis is determined by comparing the $TS$ from the real data with the distribution of $TS$ from the null hypothesis (events scrambled in r.a.).
We define the p-value as the fraction of randomized data sets with higher test statistic values than the real data.
We evaluate the median sensitivity and upper limits at a 90\% confidence level (CL) using the method of \citet{Feldman:1997qc} and calculate the discovery potential as the flux required for 50\% of trials with simulated signal to yield a p-value less than $2.87\times10^{-7}$ (i.e. $5\sigma$ significance if expressed as the one-sided tail of a Gaussian distribution).
The distribution of $TS$ for 10 million trials is shown in Fig.~\ref{fig:loglambda} for a fixed point source at $\delta=45^{\circ}$.  
Distributions with simulated $E^{-2}$ signal events injected are included, as well as a $\chi^2$ distribution with 2 degrees of freedom, which is used to estimate the $5\sigma$ significance threshold for calculating the discovery potential since simulating enough scrambled data sets requires a large amount of processing time.

\begin{figure} [ht!]
\epsscale{1.0}
\plottwo{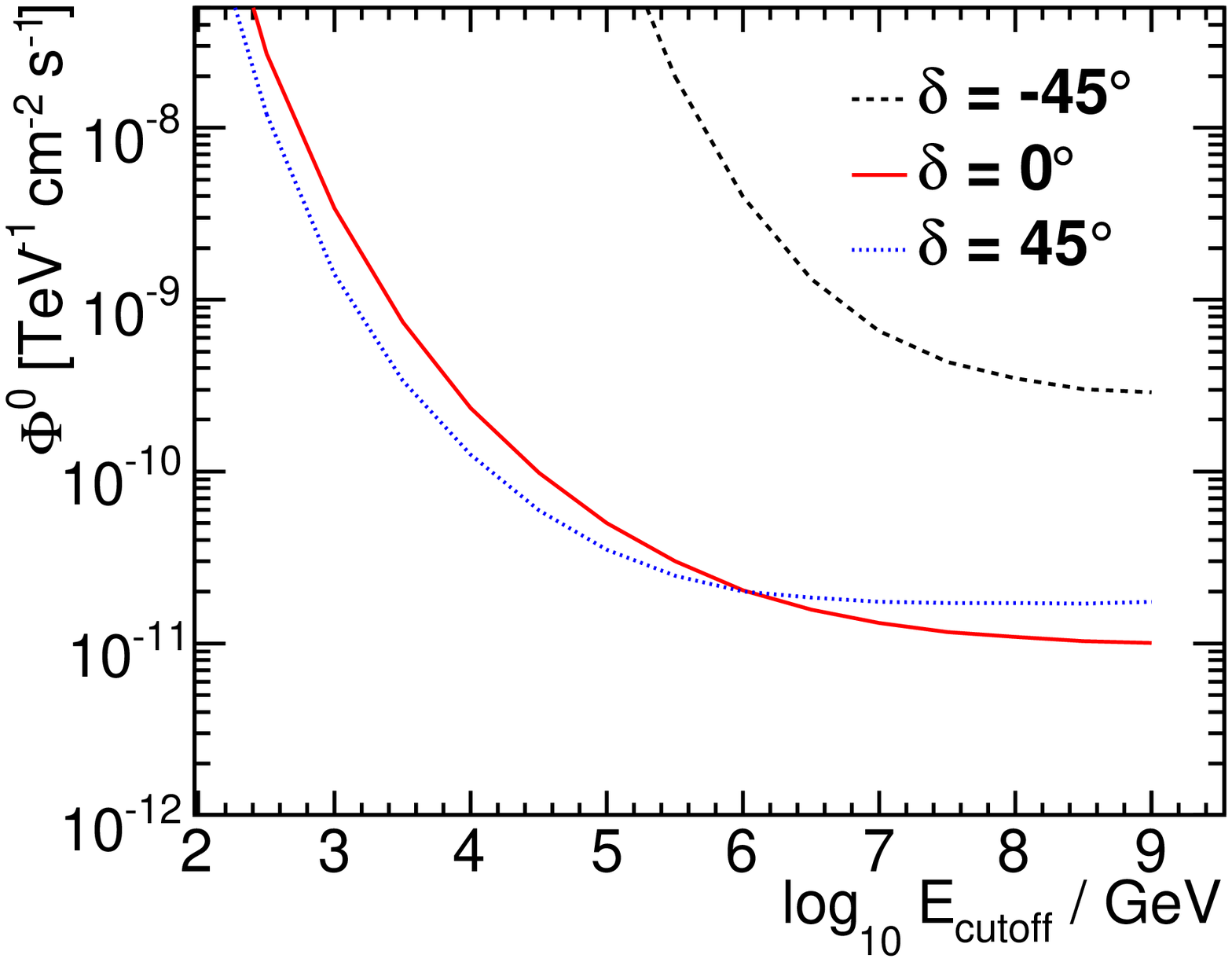}{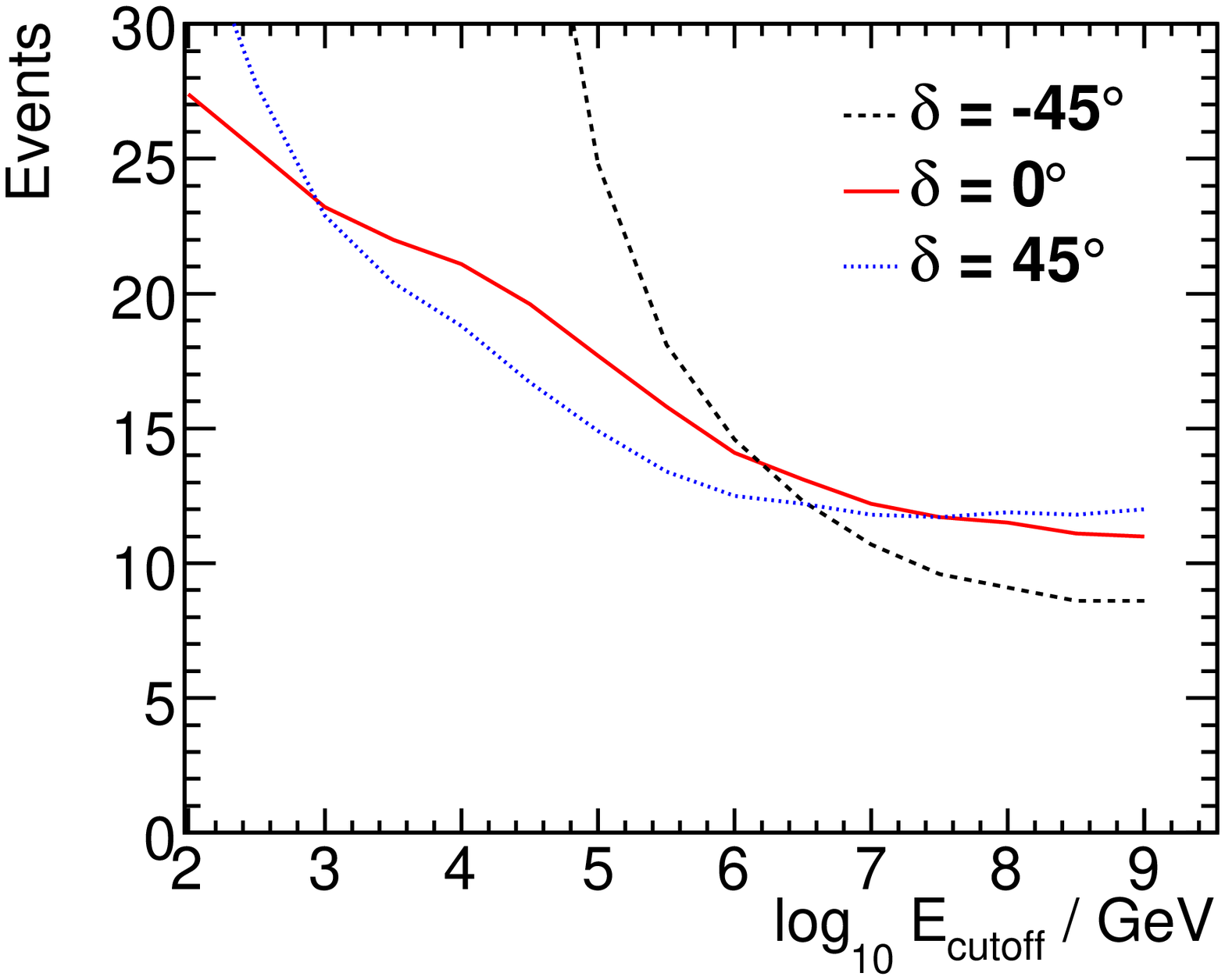}
\caption{\label{fig:cutoff}  Flux normalization (left) and number of events at the final level (right) for discovery potential versus an exponential cutoff for a differential flux parametrized as $dN/dE = \Phi^0 \cdot (E / \mathrm{TeV})^{-2}  \mathrm{exp}(-E/E_{\mathrm{cutoff}})$.  Curves are shown at three representative declinations.  The likelihood fit is still performed using a pure power law.}
\end{figure}

Although $E^{-2}$ sensitivities and limits have become a useful benchmark for comparing performance, a wide range of other spectral indices are possible along with cutoffs over a wide range of energy.  
To understand the ability of the method to detect sources with cut-off spectra, typically observed in gamma rays to be in the range 1--10~TeV for galactic sources, 
Fig.~\ref{fig:cutoff} shows the discovery potentials for a wide range of exponential cut-offs, demonstrating the ability of the method to detect sources with cut-off spectra.  
Typically, cut-offs observed in gamma rays are in the range 1--10~TeV for galactic sources.
The likelihood fit is still performed using a pure power law.  

\begin{figure} [ht!]
\epsscale{.75}
\plotone{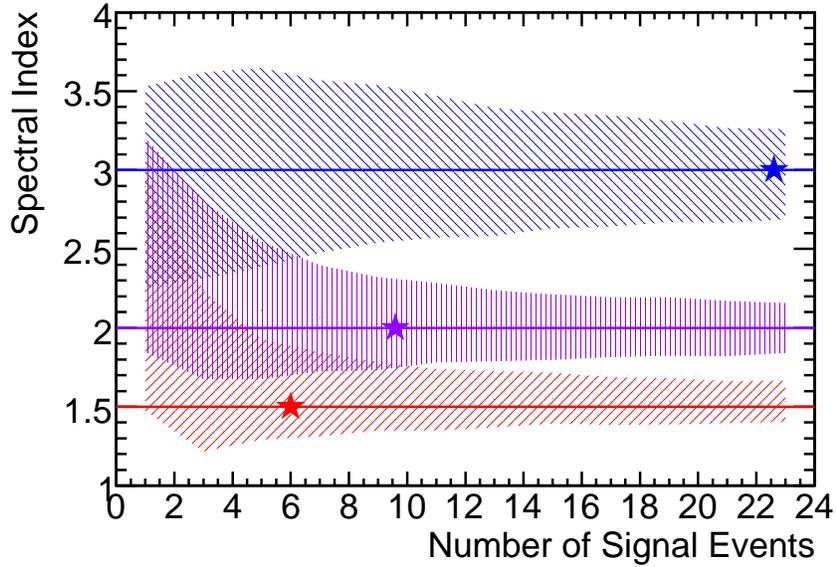}
\caption{\label{fig:spectralindex}  Reconstructed spectral index (1$\sigma$ shaded area) versus the number of signal events injected for three source spectra: $E^{-1.5}$, $E^{-2}$, and $E^{-3}$.  The sources are pure power-laws at a declination of 6$^{\circ}$.  The stars mark the number of events required for a 5$\sigma$ discovery for each spectrum.  Systematic errors are not included.}
\end{figure}

The likelihood analysis is not only more sensitive than binned methods, but it can also help extract astrophysical information.
Fig.~\ref{fig:spectralindex} shows our ability to reconstruct the spectral index for power law neutrino sources at a declination of $6^{\circ}$.  
The effective area is high for a broad range of energies here, and the spectral resolution is best.  
For each spectrum shown, the statistical uncertainty ($1\sigma$ CL) in the spectral index will be about $\pm0.3$ when enough events are present to claim a discovery.  
Spectral resolution worsens for sources farther from the horizon, to $\pm0.4$ at both $\delta=-45^{\circ}$ and $\delta=45^{\circ}$ when enough events are present for a discovery in each case.  

\begin{figure} [ht!]
\epsscale{.75}
\plotone{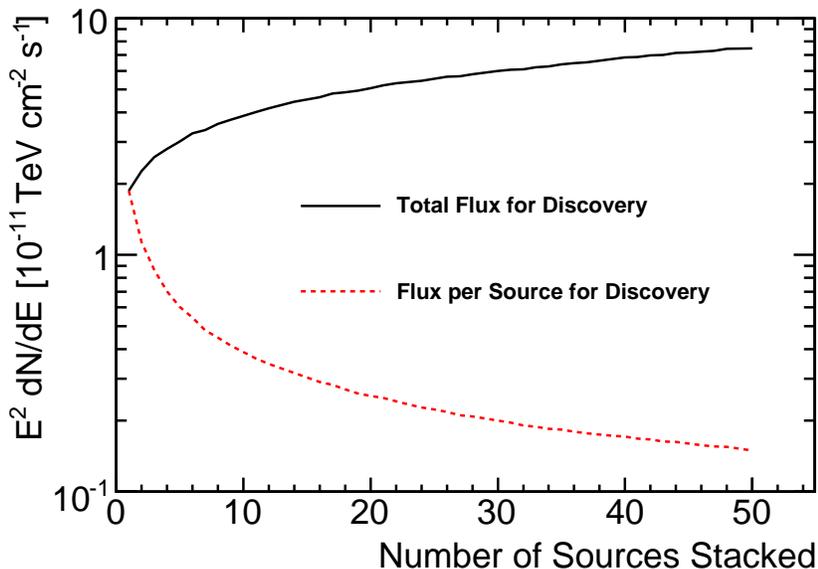}
\caption{\label{fig:nsources_stacked}  
Discovery potential flux versus the number of stacked sources, all with an equal $E^{-2}$ flux and (for example purposes) at a fixed declination of $45^{\circ}$. 
Although the total flux required increases as more sources are added, owing to a higher effective background, the flux per source required can be substantially reduced.  
}
\end{figure}

Stacking multiple sources in neutrino astronomy has been an effective way to enhance discovery potential and further constrain astrophysical models \citep{Achterberg:2006ik, AMANDA7YR_Collaboration:2008ih}.
We can consider the accumulated signal from a collection of sources using a method similar to \citet{Abbasi:2005qy}.  
Only a modification to the signal likelihood is necessary in order to stack sources, breaking the signal hypothesis into the sum over M sources:
\begin{equation}
\mathcal{S}_i \Rightarrow \mathcal{S}_i^{tot} = \frac{\sum_{j=1}^{M} W^j R^j(\gamma) \mathcal{S}_i^j}{\sum_{j=1}^{M} W^j R^j(\gamma) },
\end{equation}
where $W^j$ is the relative theoretical weight, $R^j(\gamma)$ is the relative detector acceptance for a source with spectral index $\gamma$ (assumed to be the same for all stacked sources), and $\mathcal{S}_i^j$ is the signal probability density for the $i^{th}$ event, all for the $j^{th}$ source.  
As before, the total signal events $n_s$ and collective spectral index $\gamma$ are fit parameters.  
The $W^j$ coefficients depend on our prior theoretical assumptions about the expected neutrino luminosity.  
They are higher for sources that are, on theoretical grounds, expected to be brighter.  
Tables for $R^j(\gamma)$, given as the mean number of events from a source with $dN/dE \propto E^{-\gamma}$, are calculated using simulation.
The flexibility built into the method by the relative detector acceptance and theoretical weights allows us to use source catalogs covering the whole sky and with large variations in source strengths, as well as to directly test model predictions.  
The improvement in discovery potential from stacking sources is illustrated in Fig.~\ref{fig:nsources_stacked}, showing the discovery potential flux versus the number of sources stacked.  
All sources are at the same declination in this example.

\begin{figure} [ht!]
\epsscale{.75}
\plotone{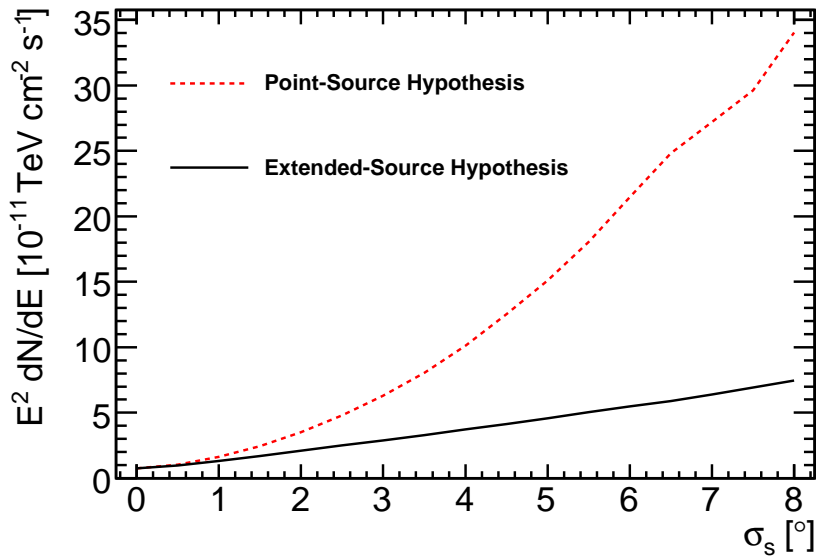}
\caption{\label{fig:extsource}  
Discovery potential flux versus the $\sigma_s$ of an extended source (distributed as a 2-dimensional Gaussian) with an $E^{-2}$ spectrum at a declination of $25^{\circ}$. 
The case of a point-source hypothesis is compared against the correct extended-source hypothesis matching what was used to simulate the signal.
}
\end{figure}

We would also like to consider sources that are spatially extended (with respect to the point spread function).  
For an example of how important this can be, the significance observed by the Milagro experiment in the location of the Fermi source J0634.0+1745 (associated with the Geminga pulsar) rises from $3.5\sigma$ to $6.3\sigma$ by fitting for an extended source \citep{MilagroFermi_Abdo:2009ku}.  
The only modification to the method required is to convolve the source distribution with the point spread function.  
Since we model our point spread function as a circular 2-dimensional Gaussian distribution, it is easy to also model a source as a circular 2-dimensional Gaussian of width $\sigma_\mathrm{s}$.  
The convolution results in a broader 2-dimensional Gaussian of width $\sqrt{\sigma_i^2 + \sigma_{\mathrm{s}}^2}$ and the likelihood uses this distribution for the signal spatial term.  
The discovery potential flux for a range of source extensions is shown in Fig.~\ref{fig:extsource} and compared to the (incorrect) hypothesis of a point source.   
For a source with true extent $\sigma_s=2^{\circ}$, the point-source hypothesis requires nearly a factor of 2 times more flux for discovery compared to the correct extended-source hypothesis.  

\section{Description of the Five Searches}
\label{sec5}

We have performed five searches:
\begin{enumerate}
\item a scan for the most significant point source in the entire sky;
\item a search over an {\em a priori} defined list of 39 interesting astrophysical objects;
\item a stacking search for 16 Milagro TeV gamma ray sources, some seen only in coincidence with the Fermi-LAT, and one unconfirmed hot spot (17 total sources);
\item a stacking search for 127 local starburst galaxies \citep{Becker:2009hw};
\item a stacking search for five nearby clusters of galaxies (CGs), testing four different models for the CR spatial distribution \citep{Murase:2008yt}.
\end{enumerate}

%
The analyses and event selection procedure were determined before unblinding the r.a. of the data.
We require a $5\sigma$ significance for discovery.  
Final p-values are calculated for each search individually.  

\subsection{All-sky Scan}
\label{sub:scan}

The first search is a scan for the single most significant point source of neutrinos over the declination range $-85^{\circ}$ to $+85^{\circ}$.  
The maximum likelihood ratio is defined continuously over the sky, and we sample it on a grid of $0.1^\circ$ in r.a. and $0.1^\circ$ in dec.
The size of the grid is not important as long as it is small compared to the angular resolution of the detector.  
Using a finer grid increases the computation time with no added benefit.  
A grid size that is comparable to or larger than the angular resolution could miss the location of the peaks in the significance map, yielding sub-optimal performance.  

\subsection{{\em A Priori} Source List}
\label{sub:point}

In order to avoid the large number of effective trials associated with scanning the entire sky, we also perform a search for the most significant of 39 {\it a priori} selected source candidates, given in Table~\ref{tab:sourcelist}.  
These sources have been selected on the basis of observations in gamma rays or astrophysical modeling that predicts neutrino emission. 
We also added the most significant location observed in the 22-string IceCube configuration (a post-trial p-value of 1.3\%  \citep{Abbasi:2009iv}).

\subsection{Milagro TeV Source Stacking}
\label{sub:stack1}

The Milagro Collaboration has reported 16 sources of TeV gamma rays \citep{Milagro_Abdo:2007ad}, several only after correlating with GeV gamma rays from the Fermi Gamma-ray Space Telescope source list \citep{MilagroFermi_Abdo:2009ku}.  
These sources are promising candidates for detection by neutrino telescopes.   Particularly interesting are sources in the complex Cygnus region \citep{Beacom:2007yu} and six supernova remnant associations \citep{Halzen_Kappes_OMurchadha_Halzen:2008zj, GonzalezGarcia:2009jc}, including MGRO J1852+01, a hot spot that falls below the significance threshold of the Milagro Coll. to be claimed as a source.  
If confirmed as a source, MGRO J1852+01 could contribute a large fraction (about 42\%) of the total neutrino flux from the SNR sources \citep{Halzen_Kappes_OMurchadha_Halzen:2008zj}.  
For the 40-string configuration of IceCube, the model of \citet{Halzen_Kappes_OMurchadha_Halzen:2008zj} predicts 3.0 neutrino events in 375.5 days, following an $E^{-2.1}$ spectrum with an exponential cutoff at about 600~TeV.  

We performed a stacking search for 17 sources observed in TeV gamma rays by Milagro (adding MGRO J1852+01 to the 16 sources which were found significant by the Milagro Coll.) using an equal weight for each source in the likelihood.  
Assuming that neutrino and gamma ray fluxes correlate and using these as weights in the likelihood did not appreciably improve the sensitivity in this case.  
Spatial extensions were used in the search for three of the sources where measurements were given (also used in the source simulation for limit calculations).  
The largest source was MGRO J2031+41, reported to have a diameter of $3.0^{\circ} \pm 0.9^{\circ}$.  

\subsection{Starburst Galaxy Stacking}
\label{sub:stack2}

Starburst galaxies have a dense interstellar medium and high star formation rates, particularly of high-mass stars.  
This leads to both high supernova rates and heating of ambient dust.  
The model of \citet{Becker:2009hw} associates the far infrared (FIR) emission with this hot dust and the radio emission with synchrotron losses of CR electrons, presumably accelerated along with hadronic CRs in the elevated number of SNRs.  
The observed strong correlation between the FIR and radio emission points to the high star formation rate as the single underlying cause, and should also correlate with the neutrino flux.   
The increased production of CRs and high density of target material are ideal conditions for neutrino production.
The starburst galaxies M82 and NGC\,253 have been observed in gamma rays at GeV--TeV energies \citep{Abdo:2009as, VERITAS_M82, Aharonian:2005de} and are the only observed steady extragalactic TeV gamma ray sources not associated with AGNs.  

We performed a stacking search for 127 starburst galaxies, weighting the sources by their observed FIR flux at $60~\mu$m, as compiled in Table A.1 in \citet{Becker:2009hw}.

\subsection{Galaxy Cluster Stacking}
\label{sub:stack3}

Clusters of galaxies (CGs) are another potential source of high energy protons and, through interactions with intracluster material (ICM), neutrinos.  
CGs are the largest gravitationally bound objects in the Universe and continue to grow through merging and accretion of dark matter and baryonic gas, generating shock fronts on megaparsec scales.  
The possibility for CGs to be sources of ultra high energy CRs above $3\times10^{18}$ eV is described in, e.g., \citet{Norman:1995,Kang:1996rp}.  
\citet{Murase:2008yt} discuss the possibility of CGs being a significant contribution to the CR spectrum between the second knee at about $3\times10^{17}$ eV and the ankle at about $3\times10^{18}$ eV.  
They give predictions for neutrinos from five nearby ($z<0.03$) CGs: Virgo, Centaurus, Perseus, Coma, and Ophiuchus.  
Information on location, distance, and size of CGs (virial radii) was taken from \citet{ReiprichClusterProps}.
These nearby CGs appear to us as spatially extended objects with virial radii subtending 1.3$^{\circ}$--6.9$^{\circ}$, so an extended spatial distribution of neutrinos is possible.  
Whereas the distribution of the ICM is well known from X-ray observations \citep{Pfrommer:2003mk}, the distribution of CRs is highly uncertain.  
The distribution of neutrinos is given by the product of the CR and ICM distributions.  
Four CR models have been considered for neutrino production, discussed in \citet{Murase:2008yt} and references therein (e.g., \citep{Colafrancesco:1998us, Berezinsky:1996wx}):

\begin{itemize}
\item{
{\bf Model A:} CRs are uniformly distributed within the cluster shock radius, taken to be 0.56 of the virial radius for the dynamical parameters considered.
}
\item{
{\bf Model B:} CRs are uniformly distributed within the virial radius, yielding the most conservative neutrino flux distributed over the largest area.
}
\item{
{\bf Isobaric:} CRs follow the distribution of thermal gas.
}
\item{
{\bf Central AGN:} In a two-step acceleration scenario CRs are accelerated in the central AGN up to a maximum energy before diffusing throughout the cluster and possibly undergoing further acceleration.  For the purposes of IceCube searches, this model can be treated as a point source.  This model is discussed in detail by \citet{Kotera:2009ms}.  
}
\end{itemize}

Signal neutrinos were simulated according to each of the four models.  
We modeled the source extensions in the likelihood as 2-dimensional Gaussian distributions with the width for each source and each model determined by optimizing for the best discovery potential.  
Although the modeling of the source extension as a Gaussian in the likelihood is not ideal, it is straightforward and computationally fast.  
The exact shape of the sources is not important for small signals; we may be able to analyze the shape with more detail depending on the intensity of any signal.  

We performed a stacking search for five nearby CGs mentioned above following the model predictions of \citet{Murase:2008yt} as weights in the likelihood.  
The size of the clusters in the likelihood fit was allowed to vary discretely between the optimal widths for each CR distribution model.  
The optimal width and $\nu_{\mu}$ differential flux for each source and each model are given in Table~\ref{tab:cluster_params}.  
The differential fluxes are parametrized as broken power laws:  
\begin{equation}
\frac{dN}{dE}~[\mathrm{TeV}^{-1} \mathrm{cm}^{-2} \mathrm{s}^{-1}] = \left\{
     \begin{array}{lr}
       A \cdot (E / \mathrm{TeV})^{-\gamma_1} & : E \leq E_{\mathrm{break}}\\
       B \cdot (E / \mathrm{TeV})^{-\gamma_2} & : E > E_{\mathrm{break}}
     \end{array}
   \right.
\label{eq:broken_power}
\end{equation}
The parameter $B = A \cdot E_{\mathrm{break}}^{\gamma_2-\gamma_1}$ after enforcing continuity at the break energy.

\begin{deluxetable}{c c c | c c c c c c }
\tabletypesize{\scriptsize}
\tablewidth{0pt}
\tablecaption{\label{tab:cluster_params} Galaxy cluster parameters.}
\startdata
\hline \hline
Source & r.a.~[$^{\circ}$] & dec.~[$^{\circ}$] & Model & $\sigma_s$~[$^{\circ}$] & $A$~[TeV$^{-1}$ cm$^{-2}$ s$^{-1}$] & $\gamma_1 $ & $\gamma_2$ & $E_{\mathrm{break}}$~[TeV] \\
\hline 
              \multirow{4}{*}{Virgo} & \multirow{4}{*}{186.63} & \multirow{4}{*}{12.72} & Model A & 2.0 & $1.42\times10^{-12}$ & -2.14 & -4.03 & $2.16\times10^{6}$ \\
              & & & Model B & 4.0 & $1.18\times10^{-12}$ & -2.14 & -4.03 & $2.16\times10^{6}$ \\
              & & & Isobaric & 3.0 & $7.57\times10^{-13}$ & -2.14 & -4.03 & $2.16\times10^{6}$ \\
              & & & Central AGN & 0.0 & $6.47\times10^{-12}$ & -2.42 & -4.24 & $2.13\times10^{6}$ \\
\hline 
               \multirow{4}{*}{Centaurus} & \multirow{4}{*}{192.20} & \multirow{4}{*}{-41.31} & Model A & 0.25 & $2.78\times10^{-13}$ & -2.14 & -4.03 & $2.15\times10^{6}$ \\
               & & & Model B & 0.5 & $2.20\times10^{-13}$ & -2.14 & -4.03 & $2.15\times10^{6}$ \\
               & & & Isobaric & 0.25 & $1.09\times10^{-13}$ & -2.15 & -4.07 & $2.33\times10^{6}$ \\
               & & & Central AGN & 0.0 & $5.10\times10^{-13}$ & -2.45 & -4.28 & $2.39\times10^{6}$ \\
\hline 
               \multirow{4}{*}{Perseus} & \multirow{4}{*}{49.95} & \multirow{4}{*}{41.52} & Model A & 0.0 & $5.83\times10^{-14}$ & -2.15 & -4.07 & $2.32\times10^{6}$ \\
               & & & Model B & 0.5 & $4.60\times10^{-14}$ & -2.15 & -4.07 & $2.32\times10^{6}$ \\
               & & & Isobaric & 0.0 & $6.17\times10^{-13}$ & -2.15 & -4.07 & $2.32\times10^{6}$ \\
               & & & Central AGN & 0.0 & $5.97\times10^{-13}$ & -2.40 & -4.20 & $1.88\times10^{6}$ \\
\hline 
               \multirow{4}{*}{Coma} & \multirow{4}{*}{194.95} & \multirow{4}{*}{27.94} & Model A & 0.25 & $2.14\times10^{-14}$ & -2.14 & -4.03 & $2.12\times10^{6}$ \\
               & & & Model B & 0.25 & $1.34\times10^{-14}$ & -2.14 & -4.03 & $2.12\times10^{6}$ \\
               & & & Isobaric & 0.25 & $1.83\times10^{-13}$ & -2.15 & -4.07 & $2.30\times10^{6}$ \\
               & & & Central AGN & 0.0 & $2.13\times10^{-13}$ & -2.41 & -4.20 & $1.89\times10^{6}$ \\
\hline 
               \multirow{4}{*}{Ophiuchus} & \multirow{4}{*}{258.11} & \multirow{4}{*}{-23.36} & Model A & 0.0 & $4.87\times10^{-14}$ & -2.15 & -4.07 & $2.29\times10^{6}$ \\
               & & & Model B & 0.5 & $1.50\times10^{-14}$ & -2.15 & -4.07 & $2.29\times10^{6}$ \\
               & & & Isobaric & 0.0 & $5.50\times10^{-13}$ & -2.15 & -4.11 & $2.49\times10^{6}$ \\
               & & & Central AGN & 0.0 & $2.55\times10^{-13}$ & -2.43 & -4.24 & $2.12\times10^{6}$ \\

\enddata
\tablecomments{
$\sigma_s$ is the optimized sigma of a 2-dimensional Gaussian distribution used in the likelihood.  Numerical calculations of the differential fluxes \citep{Murase:2009personal} for each model described in \citet{Murase:2008yt} are fit well to broken power laws, parametrized in Eq.~\ref{eq:broken_power}. }
\end{deluxetable}

\section{Systematic Errors}
\label{sec6}


The main systematic uncertainties on the flux limits come from photon propagation in ice, absolute DOM sensitivity, and uncertainties in the Earth density profile as well as muon energy loss.  
All numbers are for an $E^{-2}$ neutrino spectrum of muon neutrinos.  
We evaluate the systematic uncertainty due to photon propagation by performing dedicated simulations with scattering and absorption coefficients varied within their uncertainties of $\pm10\%$ \citep{OpticalProperties_JGeophysRes_2006}.  
The maximum difference was between the case where both scattering and absorption were increased by 10\% and the case where both were decreased by 10\%.  
The deviation in the observed number of events between these two cases was 11\%.  
The range of uncertainty in the DOM sensitivity is taken as $\pm 8\%$, based on the measured uncertainty in the PMT sensitivity \citep{Abbasi:2010vc}.
Similarly, another dedicated simulation where we varied the DOM sensitivity inside the uncertainty leads to a maximum systematic uncertainty on the number of events of $9\%$.
These uncertainties on the flux varied by only about 2\% between the northern and southern sky, so only averages over the whole sky are reported.  
Finally, uncertainties in muon energy losses, the neutrino-nucleon cross-section, and the rock density near the detector introduce an 8\% systematic uncertainty for vertically up-going events \citep{AMANDA5YR_Achterberg:2006vc}.  
These events are the most affected, and this uncertainty is applied to all zeniths to be conservative.  
A sum in quadrature of the systematic uncertainties on the flux gives a total of 16\% systematic uncertainty in the signal simulation.   
These systematic uncertainties are incorporated into the upper limit and sensitivity calculations using the method of \citet{Conrad:2002kn} with a modification by \citet{Hill:2003jk}.

The analyses described in Sec.~\ref{sec5} give reliable statistical results (p-values) due to the ability to generate background-only data sets by scrambling the data in right ascension.  
By using the data to estimate background, the systematic errors come only from signal and detector simulation used to calculate flux upper limits, sensitivities, and discovery potentials.
Nevertheless, the comparison of background simulation (both atmospheric neutrinos and atmospheric muons) with data is an important check to build confidence in the signal simulation.  

\begin{figure} [ht!]
\epsscale{.75}
\plotone{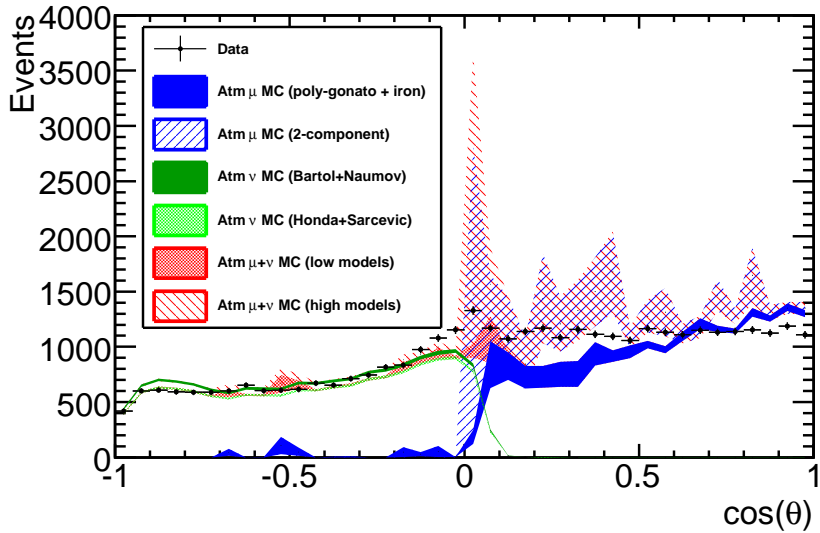}
\caption{\label{fig:final_zenith}  
Distribution of reconstructed cosine zenith for the final event sample compared to the models discussed in the text.  Honda and Sarcevic are summed with {\em poly-gonato} to represent the set of low predictions, and Bartol and Naumov are summed with the 2-component model for the high predictions.  Only statistical errors are shown. The two-component model has limited statistics, causing the peaks and valleys.  Systematic uncertainties of neutrino production in CR showers are estimated to be about 40\% at 1~TeV \citep{BartolUncertainty_Barr:2006it} and 15\% in the muon rate greater than about 1~TeV \citep{Ahn:2009wx,Berghaus:2008zz}.}
\end{figure}

Muons created in CR showers in the atmosphere at zenith angles around 80$^{\circ}$ (near the horizon) must travel about 15~km to reach the bottom of IceCube.  
Only very high energy muons can travel such distances.
For the simulation to produce the correct zenith distribution for these nearly horizontal events, CR composition can be important since protons can produce higher energy muons than iron nuclei with the same energy.  
Two CR models were simulated: 
First, a slightly modified version of the {\em poly-gonato} model of the galactic CR flux and composition \citep{Hoerandel:2002yg}.  
Above the galactic model cutoff at $Z \times 4\times10^{15}$~eV, a flux of pure iron is used with an $E^{-3}$ spectrum.  
As discussed in \ref{sub:simu}, these corrections are required to reproduce the measured CR spectrum at these energies.  
Second, a simpler pure proton and iron two-component model with a much higher contribution of protons is also used for comparison \citep{Glasstetter1999238}.  
The final zenith distribution with each of these models is shown in Fig.~\ref{fig:final_zenith}.  
The atmospheric muon simulation is not only affected by the primary composition uncertainties at high energy; it is also affected by the hadronic model, affecting the production rate of muons at the level of 15\% in the region of interest for IceCube, greater than about 1~TeV, as discussed in the SIBYLL model paper \citep{Ahn:2009wx} and in the comparison between different hadronic models used in CORSIKA presented in \citet{Berghaus:2008zz}.
For the up-going region, several models of atmospheric neutrino fluxes, both conventional fluxes from pion and kaon decay and prompt fluxes from charmed meson decay, are shown in Fig.~\ref{fig:final_zenith}.  
To represent the low and high predictions, conventional and prompt models are used in pairs: Honda \citep{Honda:2006qj} for the conventional flux paired with Sarcevic \citep{Enberg:2008te} for the prompt flux represent the low prediction, and Bartol \citep{Barr:2004br} for the conventional flux paired with Naumov \citep{Naumov} for the prompt flux represent the high prediction.  
Additional uncertainty in the predicted atmospheric neutrino rate is estimated to be about 40\% at 1~TeV \citep{BartolUncertainty_Barr:2006it}.  
We conclude that our data agree with background simulation at the final level, within the range of uncertainties allowable by existing CR composition and atmospheric neutrino models.  

\section{Results}
\label{sec7}

\begin{figure}
\epsscale{1.0}
\plotone{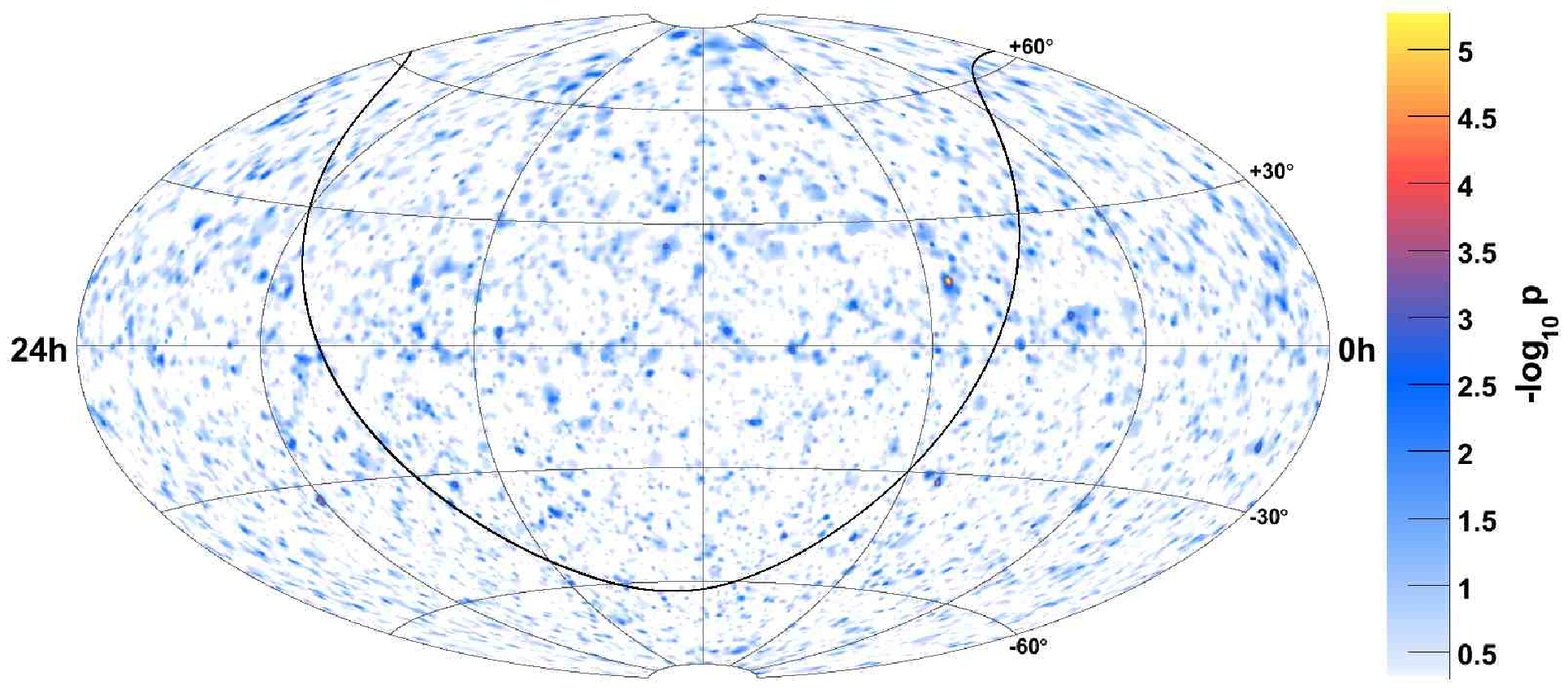}
\caption{\label{fig:skymap}Equatorial skymap (J2000) of pre-trial significances (p-value) of the all-sky point source scan.  The galactic plane is shown as the solid black curve.}
\end{figure}

\begin{figure}
\epsscale{1.0}
\plotone{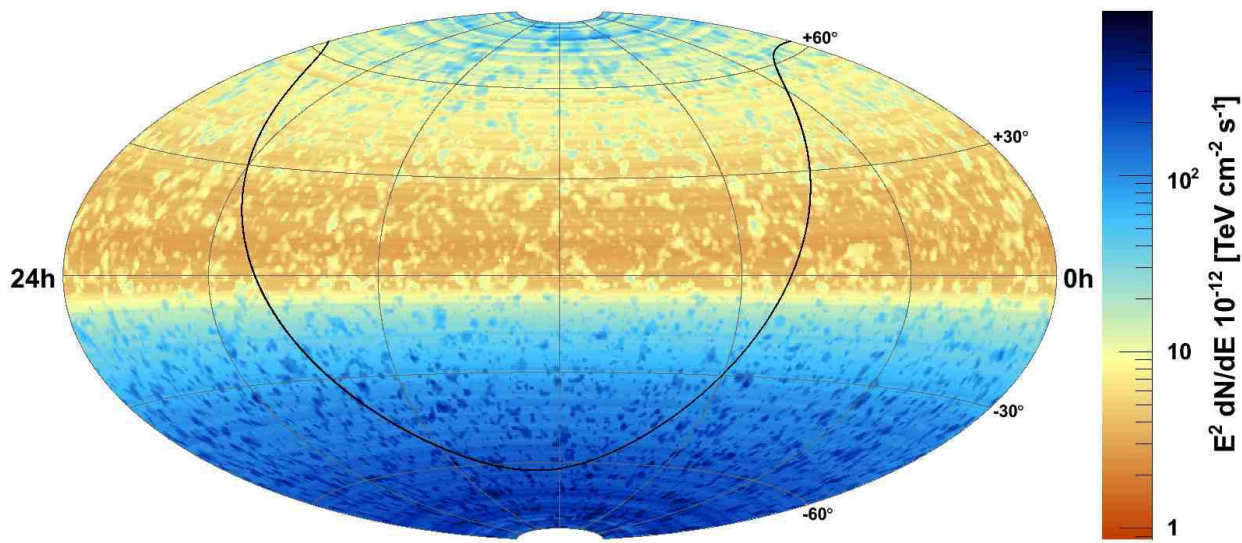}
\caption{\label{fig:uppermap}Equatorial skymap (J2000) of upper limits of Feldman-Cousins 90\% confidence intervals for an $E^{-2}$ flux of $\nu_{\mu} +\bar\nu_{\mu}$.  The galactic plane is shown as the solid black curve.}
\end{figure}

The results of the all-sky scan are shown in the map of the pre-trial p-values in Fig.~\ref{fig:skymap}.  
The most significant deviation from background is located at 113.75$^{\circ}$ r.a., 15.15$^{\circ}$ dec.  
The best-fit parameters are $n_s= 11.0$ signal events above background, with spectral index $\gamma=2.1$.  
The pre-trial estimated p-value of the maximum log likelihood ratio at this location is $5.2\times10^{-6}$.  
In trials using data sets scrambled in right ascension, 1,817 out of 10,000 have an equal or higher significance somewhere in the sky, resulting in the post-trial p-value of 18\%.  
Upper limits for an $E^{-2}$ flux of $\nu_{\mu} + \bar\nu_{\mu}$ are presented in Fig.~\ref{fig:uppermap}.  
In all cases, an equal flux of neutrinos and anti-neutrinos is assumed.

\begin{figure} [ht!]
\epsscale{.75}
\plotone{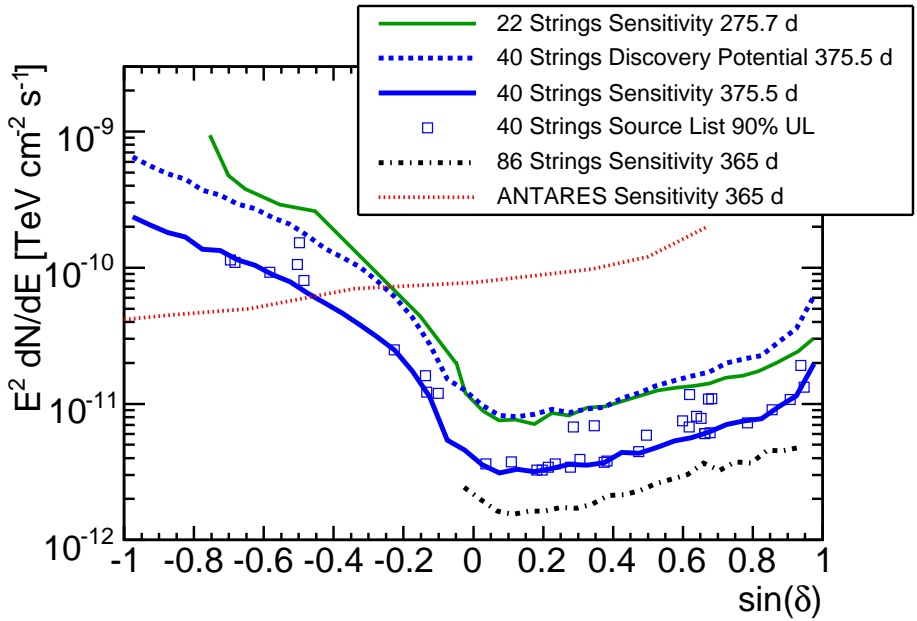}
\caption{\label{fig:sens} Median sensitivity to a point-source $E^{-2}$ flux of $\nu_{\mu} +\bar\nu_{\mu}$ as a function of declination, shown for the 22-string IceCube southern and northern sky analyses \citep{Abbasi:2009cv, Abbasi:2009iv}, this 40-string analysis, and preliminary estimated sensitivities for 1 year for the ANTARES experiment, primarily sensitive in the GeV--TeV energy range, \citep{Coyle:2010zm} and the final IceCube configuration (using the event selection based on this work for the up-going region).  For the {\em a priori} source list, upper limits of Feldman-Cousins 90\% confidence intervals for an $E^{-2}$ flux of $\nu_{\mu} +\bar\nu_{\mu}$ are shown.  In addition, we show the discovery potential for this work.}
\end{figure}

\begin{figure} [ht!]
\epsscale{.75}
\plotone{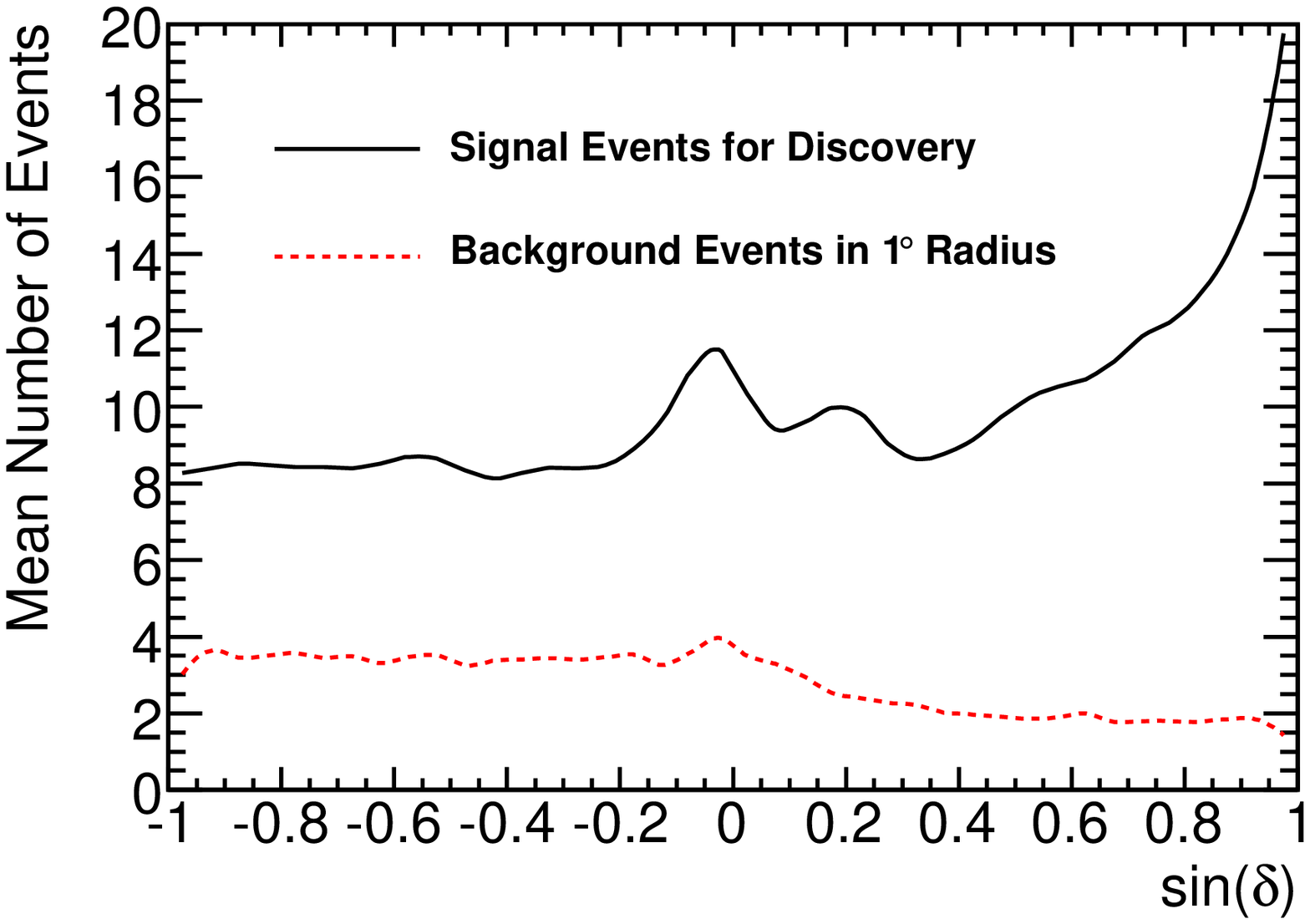}
\caption{\label{fig:events} Mean number of $E^{-2}$ $\nu_{\mu} + \bar\nu_{\mu}$ signal events at the final cut level required for a discovery at $5\sigma$ in 50\% of trials and the mean number of background events in a circular bin with a radius of $1^{\circ}$ versus sine of the declination.}
\end{figure}

The results of the point-source search in the direction of 39 source candidates selected {\em a priori} are given in Table~\ref{tab:sourcelist} and also shown in Fig.~\ref{fig:sens} with the IceCube median sensitivity.  
The most significant source on the list was PKS 1622-297 with a pre-trial estimated p-value of 5\%.  
The post-trial p-value of 62\% was again determined as the fraction of scrambled data sets with at least one source with an equal or higher significance.  
The mean number of events at the final cut level required for the discovery of a point source is also shown in Fig.~\ref{fig:events}, along with the average background in a circular bin with $1^{\circ}$ radius.  
Included in Fig.~\ref{fig:sens} is a preliminary comparison to the ANTARES experiment \citep{Coyle:2010zm}.  
ANTARES is primarily sensitive to GeV--TeV energy neutrinos in the southern sky, so the coverage in energy is quite complementary to this IceCube analysis.  



\begin{deluxetable}{c c c c c c c c c c}
\tablewidth{0pt}
\tablecaption{Results for the {\it a priori} source candidate list. \label{tab:sourcelist}}
\tablehead{
\colhead{Object} & \colhead{r.a.~[$^{\circ}$]} & \colhead{dec.~[$^{\circ}$]} & \colhead{$\Phi_{\nu_{\mu}}^{90}$} & \colhead{$\Phi_{\nu_{\mu}+\nu_{\tau}}^{90}$} & \colhead{p-value}  & \colhead{n$_s$} & \colhead{$\gamma$} & \colhead{N$_{1^{\circ}}$} & \colhead{B$_{1^{\circ}}$}
}
\startdata
             Cyg OB2 & 308.08 & 41.51 & 6.04 & 10.54 & -- & 0.0 & -- &   2 & 1.8\\
       MGRO J2019+37 & 305.22 & 36.83 & 7.50 &  13.3 & 0.44 & 1.0 & 2.8 &   2 & 1.9\\
       MGRO J1908+06 & 286.98 &  6.27 & 3.73 &  6.82 & 0.43 & 1.5 & 3.9 &   4 & 3.1\\
               Cas A & 350.85 & 58.81 & 9.04 &  15.92 & -- & 0.0 & -- &   1 & 1.8\\
               IC443 &  94.18 & 22.53 & 3.80 &  6.62 & -- & 0.0 & -- &   1 & 2.0\\
             Geminga &  98.48 & 17.77 & 3.91 &  6.66 & 0.48 & 0.7 & 2.1 &   1 & 2.3\\
         Crab Nebula &  83.63 & 22.01 & 3.70 &  6.58 & -- & 0.0 & -- &   1 & 2.0\\
        1ES 1959+650 & 300.00 & 65.15 & 10.74 &  19.18 & -- & 0.0 & -- &   0 & 2.0\\
        1ES 2344+514 & 356.77 & 51.70 & 7.24 &  12.96 & -- & 0.0 & -- &   0 & 1.8\\
               3C66A &  35.67 & 43.04 & 10.89 &  19.70 & 0.24 & 3.4 & 3.9 &   3 & 1.9\\
          H 1426+428 & 217.14 & 42.67 & 6.14 &  10.94 & -- & 0.0 & --  &   3 & 1.8\\
              BL Lac & 330.68 & 42.28 & 10.80 & 18.70 & 0.25 & 2.6  & 3.9 &   3 & 1.8\\
             Mrk 501 & 253.47 & 39.76 & 8.11 &  14.14 & 0.41 & 1.3  & 3.9 &   3 & 2.0\\
             Mrk 421 & 166.11 & 38.21 & 11.71 & 20.14 & 0.15 & 2.6 & 1.9 &   2 & 2.0\\
             W Comae & 185.38 & 28.23 & 4.46 &  8.06 & -- & 0.0 & -- &   0 & 1.9\\
        1ES 0229+200 &  38.20 & 20.29 & 6.89 & 12.06 & 0.19 & 4.0 & 2.8 &   4 & 2.1\\
                 M87 & 187.71 & 12.39 & 3.42 & 5.98 & -- & 0.0 & -- &   2 & 2.5\\
          S5 0716+71 & 110.47 & 71.34 & 13.28 & 23.56 & -- & 0.0 & -- &   0 & 1.6\\
                 M82 & 148.97 & 69.68 & 19.14 & 32.84 & 0.4 & 2.0 & 3.9 &   4 & 1.8\\
            3C 123.0 &  69.27 & 29.67 & 5.59 &  10.66 & 0.44 & 1.3 & 2.7 &   1 & 1.9\\
            3C 454.3 & 343.49 & 16.15 & 3.42 & 5.92 & -- & 0.0 & -- &   1 & 2.3\\
            4C 38.41 & 248.81 & 38.13 & 6.77 & 11.86 & 0.48 & 0.9 & 3.9 &   2 & 2.0\\
        PKS 0235+164 &  39.66 & 16.62 & 6.77 & 11.62 & 0.15 & 5.3 & 3.0 &   5 & 2.3\\
        PKS 0528+134 &  82.73 & 13.53 & 3.63 & 6.72 & -- & 0.0 & -- &   2 & 2.4\\
        PKS 1502+106 & 226.10 & 10.49 & 3.26 & 5.78 & -- & 0.0 & -- &   0 & 2.5\\
              3C 273 & 187.28 &  2.05 & 3.61 &  6.54 & -- & 0.0 & -- &   3 & 3.4\\
            NGC 1275 &  49.95 & 41.51 & 6.04 & 10.54 & -- & 0.0 & -- &   2 & 1.8\\
               Cyg A & 299.87 & 40.73 & 7.84 & 13.44 & 0.46 & 1.0 & 3.5 &   3 & 1.9\\
       IC-22 maximum & 153.38 & 11.38 & 3.26 & 5.86 & -- & 0.0 & -- &   1 & 2.5\\
              Sgr A* & 266.42 & -29.01 & 80.56 & 139.26 & 0.41 & 1.1 & 2.7 &   4 & 3.3\\
        PKS 0537-441 &  84.71 & -44.09 & 113.90 & 201.82 & -- & 0.0 & -- &   3 & 3.5\\
               Cen A & 201.37 & -43.02 & 109.51 & 191.56 & -- & 0.0 & -- &   4 & 3.5\\
        PKS 1454-354 & 224.36 & -35.65 & 92.56 & 156.74 & -- & 0.0 & -- &   4 & 3.5\\
        PKS 2155-304 & 329.72 & -30.23 & 105.41 & 182.90 & 0.28 & 1.7 & 3.9 &   3 & 3.4\\
        PKS 1622-297 & 246.53 & -29.86 & 152.28 & 263.86 & 0.048 & 3.0 & 2.6 &   4 & 3.3\\
        QSO 1730-130 & 263.26 & -13.08 & 24.83 & 43.30 & -- & 0.0 & -- &   4 & 3.5\\
        PKS 1406-076 & 212.24 & -7.87 & 16.04 & 28.72 & 0.42 & 1.3 & 2.3 &   4 & 3.3\\
        QSO 2022-077 & 306.42 & -7.64 & 12.18 & 21.78 & -- & 0.0 & -- &   2 & 3.3\\
               3C279 & 194.05 & -5.79 & 11.94 & 21.36 & 0.33 & 3.6 & 3.0 &   7 & 3.5\\
\enddata
\tablecomments{
$\Phi_{\nu_{\mu}}^{90}$ and $\Phi_{\nu_{\mu}+\nu_{\tau}}^{90}$ are the upper limits of the Feldman-Cousins 90\% confidence intervals for an $E^{-2}$ flux normalization of $\nu_{\mu}$ and $\nu_{\mu}+\nu_{\tau}$, respectively, in units of $10^{-12}~\mathrm{TeV}^{-1} \mathrm{cm}^{-2} \mathrm{s}^{-1}$ (i.e. $dN/dE \leq \Phi^{90} \cdot (E/\mathrm{TeV})^{-2}$).  $n_{s}$ is the best-fit number of signal events; when $n_{s}>0$ the (pre-trial) p-value is also calculated and the spectral index $\gamma$ is given.  N$_{1^{\circ}}$ is the actual number of events observed in a bin of radius $1^{\circ}$. The background event density at the source declination is indicated by the mean number of background events $B_{1^{\circ}}$ expected in a bin of radius $1^{\circ}$.}
\end{deluxetable}

The Milagro TeV source stacking search resulted in a post-trial p-value of 32\% with best fit $n_s=7.6$ and spectral index $\gamma=2.6$.  
The starburst galaxy stacking search resulted in an under-fluctuation with best fit $n_s=0$, since we do not allow negative values of $n_s$.
Finally, the CGs stacking search yielded a post-trial p-value of 78\% with $n_s=1.8$.
These results, along with sensitivities and upper limits, are summarized in Table~\ref{tab:stackingresults}.

\begin{deluxetable}{c c c c c c c c}
\tabletypesize{\scriptsize}
\rotate
\tablewidth{0pt}
\tablecaption{\label{tab:stackingresults} Results for the stacked source searches.}
\tablehead{
\colhead{Catalog} & \colhead{N Sources} & \colhead{Model} & \colhead{p-value} & \colhead{${\nu_{\mu}}$ Sensitivity} & \colhead{${\nu_{\mu}}$ Upper Limit} & \colhead{${\nu_{\mu}+\nu_{\tau}}$ Sensitivity}  & \colhead{${\nu_{\mu}+\nu_{\tau}}$ Upper Limit}
}
\startdata
              \multirow{2}{*}{Milagro Sources} & 17 & $E^{-2}$, Uniform & 0.32 & $\Phi^{90} = 9.0$ & $\Phi^{90} = 12.3$ & $\Phi^{90} = 15.8$ &  $\Phi^{90} = 24.5$\\
              & 6 & 6 SNR Assoc.\tablenotemark{a} & \tablenotemark{c} & & & SF = 2.9 & SF = 7.2 \\
             \hline
             Starburst Galaxies & 127 & $E^{-2}, \propto$ FIR Flux & -- &  $\Phi^{90} = 33.1$ &  $\Phi^{90} = 33.1$ & $\Phi^{90} = 58.6$ & $\Phi^{90} = 58.6$\\ 
             \hline
              \multirow{4}{*}{Clusters of Galaxies} & \multirow{4}{*}{5} & Model A\tablenotemark{b} &  \multirow{4}{*}{0.78} &  &  & SF = 8.4 & SF = 7.8\\
              & & Model B\tablenotemark{b} &  &  &  & SF = 14.4 & SF = 12.0 \\
              & & Isobaric\tablenotemark{b} &  &  &  & SF = 13.2 & SF = 13.2 \\
              & & Central AGN\tablenotemark{b} &  &  & & SF = 6.0 & SF = 6.0 \\
\enddata
\tablecomments{
Median sensitivities and upper limits at 90\% CL for $\nu_{\mu}$ and  ${\nu_{\mu}+\nu_{\tau}}$ fluxes are given in two ways: as $\Phi^{90}$ for an $E^{-2}$ spectrum, i.e. the total flux from all sources $dN/dE \leq \Phi^{90} \, 10^{-12}~\mathrm{TeV}^{-1} \mathrm{cm}^{-2} \mathrm{s}^{-1} (E / \mathrm{TeV})^{-2}$, or as a scaling factor (SF) relative to the models given in the footnotes.  For example, if the Central AGN model flux normalization were 6.0 times higher, we would rule it out at 90\% CL.  All models predict equal fluxes of tau and muon neutrinos.}
\tablenotetext{a}{\citet{Halzen_Kappes_OMurchadha_Halzen:2008zj}}
\tablenotetext{b}{\citet{Murase:2008yt}, see Table~\ref{tab:cluster_params}}
\tablenotetext{c}{We did not calculate an {\em a priori} p-value for just the six SNR associations discussed in \citet{Halzen_Kappes_OMurchadha_Halzen:2008zj}, since they are included in the search over all Milagro sources.  However, some of these sources were the most significant on the list.  Analyzed as a single subgroup, an {\em a posteriori} p-value of 0.02 was found with best fit parameters $n_s=15.2$ and $\gamma=2.9$.  The true trial factor is incalculable since this was done after unblinding, but these remain sources of interest for future searches.}
\end{deluxetable}


\section{Implications for Models of Astrophysical Neutrino Emission}
\label{sec8}

\begin{figure} [ht!]
\epsscale{.75}
\plotone{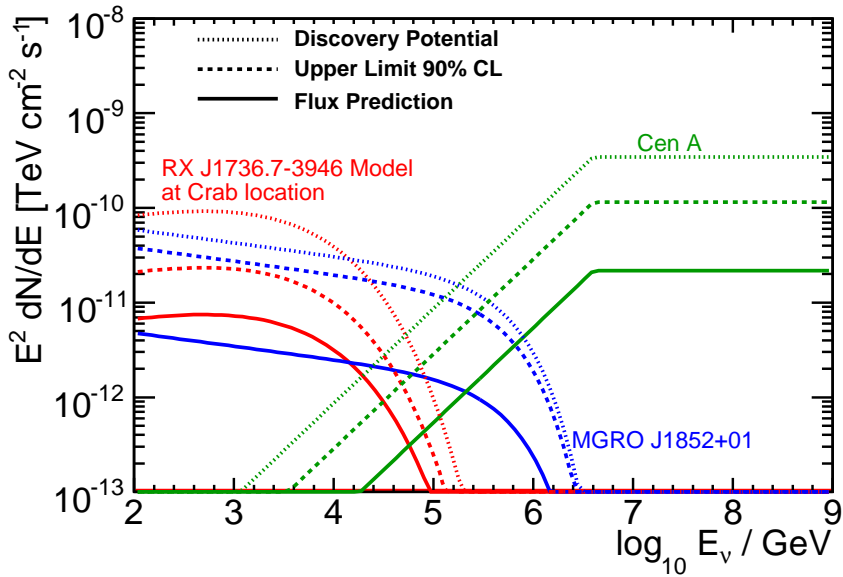}
\caption{\label{fig:models}  Differential flux of three theoretical models shown with the IceCube 40-string upper limit (90\% CL) and discovery potential in each case.  Shown are the $\nu_{\mu} + \bar\nu_{\mu}$ predictions of \citet{Morlino:2009ci} for SNR RX J1713.7-3946 but moved to the location of the Crab Nebula, \citet{Halzen_Kappes_OMurchadha_Halzen:2008zj} for MGRO J1852+01, and \citet{Koers:2008hv} for Cen A under the most optimistic condition, where the protons have a spectral index $\alpha_p=3$.
}
\end{figure}

The IceCube Neutrino Observatory aims to further our understanding of astrophysical phenomena, constraining models even in the absence of a detection.  
Fig.~\ref{fig:models} shows our sensitivity to some specific predictions.
The model of \citet{Morlino:2009ci} is for SNR RX J1713.7-3946.  
This analysis is largely insensitive to spectra which cut off below 100~TeV in the southern sky.  
Applying this emission model at the location of the Crab Nebula ($\delta = 22.01^{\circ}$) we obtain an upper limit that rules out a flux 3.2 times higher than the prediction.  

The Milagro hot spot MGRO J1852+01 is the brightest source of six SNR associations considered in \citet{Halzen_Kappes_OMurchadha_Halzen:2008zj}.  
The stacking results were already shown in Table~\ref{tab:stackingresults}.   
Our upper limit for just this one brightest source is a factor of 7.9 away from excluding this model at 90\% CL.
The best fit for MGRO J1852+01 is to 7.0 events with $\gamma=2.9$, which increases the upper limit compared to the average background-only case.

The nearest AGN, Centaurus A (Cen A), has been discussed as a potential source of ultra high energy CRs in the context of results from the Pierre Auger Observatory (PAO). 
The point source likelihood fit at the location of Cen A resulted in zero signal events in this analysis, setting an upper limit that is 5.3 times higher than the $\nu_{\mu}$ prediction by \citet{Koers:2008hv} for the most optimistic case where the protons have a spectral index $\alpha_p=3$. 

Starburst galaxies were already presented as sources of interest in Sec.~\ref{sub:stack2}.  
Recent detections \citep{Abdo:2009as, VERITAS_M82, Aharonian:2005de} of very high energy photons from the nearest luminous starburst galaxies M82 and NGC\,253, each characterized by star-forming regions with high supernova rates in the core, support the belief that the observed enhanced gamma ray emission is due to CR interactions.  
Under the assumption that the GeV--TeV photons originate from the decay of neutral pions produced when protons that are shock-accelerated by SNRs in the starburst core inelastically scatter against target hydrogen atoms with densities of the order of $10^{2}\,\mathrm{cm}^{-3}$ \citep{deCea:2009, Persic:2008}, an order-of-magnitude calculation of the resulting flux of muon neutrinos based on \citet{Kelner:2006} can be made.  
The muon neutrino upper limit from M82 is about 400 times higher than the rough prediction.  
For NGC~253 in the southern sky, the muon neutrino upper limit is about 6000 times higher than the prediction.

\section{Conclusions}
\label{sec9}

A search for sources of high energy neutrinos has been performed using data taken during 2008--09 with the 40-string configuration of the IceCube Neutrino Observatory.  
Five searches were performed: 1) a scan of the entire sky for point sources, 2) a predefined list of 39 interesting source candidates, 3) stacking 16 sources of TeV gamma rays observed by Milagro and Fermi, along with an unconfirmed hot spot (17 total sources), 4) stacking 127 starburst galaxies, and 5) stacking five nearby CGs, testing four different models for the CR distribution.  
The most significant result of the five searches came from the all-sky scan with a p-value of 18\%.  
The cumulative binomial probability of obtaining at least one result of this significance or higher in five searches is 63\%.  
This result is consistent with the null hypothesis of background only.  

The sensitivity of this search using 375.5 days of 40-string data already improves upon previous point-source searches in the TeV--PeV energy range by at least a factor of two, depending on declination.  
The searches were performed using a data set of up-going atmospheric neutrinos (northern sky) and higher energy down-going muons (southern sky) in a unified manner.  
During 2010--2011, 79 strings of IceCube are operating and detector construction should finish during the austral summer of the same years.  
The full IceCube detector should improve existing limits by at least another factor of two with one year of operation.  
Additional improvement is foreseeable in the down-going region by developing sophisticated veto techniques and at lower energies by using the new dense sub-array, DeepCore, to its fullest potential.

\acknowledgments
\section{Acknowledgements}
\label{sec10}

We would like to thank K. Murase for helpful discussions regarding clusters of galaxies.  

We acknowledge the support from the following agencies:
U.S. National Science Foundation-Office of Polar Programs,
U.S. National Science Foundation-Physics Division,
University of Wisconsin Alumni Research Foundation,
the Grid Laboratory Of Wisconsin (GLOW) grid infrastructure at the University of Wisconsin - Madison, the Open Science Grid (OSG) grid infrastructure;
U.S. Department of Energy, and National Energy Research Scientific Computing Center,
the Louisiana Optical Network Initiative (LONI) grid computing resources;
National Science and Engineering Research Council of Canada;
Swedish Research Council,
Swedish Polar Research Secretariat,
Swedish National Infrastructure for Computing (SNIC),
and Knut and Alice Wallenberg Foundation, Sweden;
German Ministry for Education and Research (BMBF),
Deutsche Forschungsgemeinschaft (DFG),
Research Department of Plasmas with Complex Interactions (Bochum), Germany;
Fund for Scientific Research (FNRS-FWO),
FWO Odysseus programme,
Flanders Institute to encourage scientific and technological research in industry (IWT),
Belgian Federal Science Policy Office (Belspo);
University of Oxford, United Kingdom;
Marsden Fund, New Zealand;
Japan Society for Promotion of Science (JSPS);
the Swiss National Science Foundation (SNSF), Switzerland;
A.~Gro{\ss} acknowledges support by the EU Marie Curie OIF Program;
J.~P.~Rodrigues acknowledges support by the Capes Foundation, Ministry of Education of Brazil.

\bibliographystyle{apj}
\bibliography{ic40ps}






\end{document}